\documentclass{imaman}

\jno{dpnxxx}
\usepackage{enumitem}
\usepackage{ulem}
\usepackage{natbib}
\usepackage{booktabs}
\usepackage{multirow}
\usepackage{hyperref}
\hypersetup{
	colorlinks=true,
	linkcolor=blue,
	filecolor=magenta,      
	urlcolor=blue,
	citecolor=black, 
}
\usepackage{url}
\usepackage{xcolor}
\usepackage[ruled]{algorithm2e}
\usepackage{verbatim}
\usepackage{mathtools}% http://ctan.org/pkg/mathtools
\usepackage{longtable}
\usepackage{algorithmic}
\usepackage{graphicx}
\graphicspath {{Paper_images/} }
\usepackage{soul}
\usepackage{xspace} 
\usepackage[makeroom]{cancel}

\usepackage{amsmath,amsthm,bm,mathrsfs}
 \newtheoremstyle{theorem}{6pt}{6pt}{\rm}{}{\sffamily}{ }{ }{}
 \theoremstyle{theorem}

 \newtheoremstyle{lemma}{6pt}{6pt}{\rm}{}{\sffamily}{ }{ }{}
 \theoremstyle{lemma}

\newtheoremstyle{case}{6pt}{6pt}{\rm}{}{\sffamily}{. }{ }{}
 \theoremstyle{case}

 \newtheoremstyle{statement}{6pt}{6pt}{\rm}{}{\sffamily}{ }{ }{}
\theoremstyle{statement}

 \newtheoremstyle{corollary}{6pt}{6pt}{\rm}{}{\sffamily}{ }{ }{}
 \theoremstyle{corollary}

  \newtheoremstyle{definition}{6pt}{6pt}{\rm}{}{\sffamily}{ }{ }{}
 \theoremstyle{definition}

\newtheoremstyle{example}{6pt}{6pt}{\rm}{}{\sffamily}{ }{ }{}
\theoremstyle{example}

\newtheoremstyle{remark}{6pt}{6pt}{\rm}{}{\sffamily}{ }{ }{}
\theoremstyle{remark}

\newtheoremstyle{approximation}{6pt}{6pt}{\rm}{}{\sffamily}{ }{ }{}
\theoremstyle{approximation}

\newtheoremstyle{scheme}{6pt}{6pt}{\rm}{}{\sffamily}{ }{ }{}
\theoremstyle{scheme}

\newtheoremstyle{Algorithm}{6pt}{6pt}{\rm}{}{\sffamily}{ }{ }{}
\theoremstyle{Algorithm}

\newtheoremstyle{Assumption}{6pt}{6pt}{\rm}{}{\sffamily}{ }{ }{}
\theoremstyle{Assumption}

\newtheoremstyle{proposition}{6pt}{6pt}{\rm}{}{\sffamily}{ }{ }{}
\theoremstyle{proposition}

\newtheoremstyle{hypo}{6pt}{6pt}{\rm}{}{\sffamily}{ }{ }{}
 \theoremstyle{hypo}

  \newtheoremstyle{Step}{6pt}{6pt}{\rm}{}{}{ }{ }{}
 \theoremstyle{Step}

\numberwithin{equation}{section}
\def\citeasnoun{\cite}

\newcommand {\dn}[1] { {\boldsymbol#1} }
\newcommand {\Stan}{ {\sf Stan}\xspace}

\begin{document}
	
	%%%%%%%%%%%%%%%%%
	%	\title{Bayesian Prediction of volleyball Sets Using the Truncated Skellam and the Ordered Multinomial Models}
	\title{Bayesian Models for Prediction of the  Set-Difference in Volleyball}
	\color{black}
	\author{ {\sc Ioannis Ntzoufras$^{\dagger}$},{\sc Vasilis  Palaskas}\footnote{This work has been completed at Athens University of Economics and Business; Current affiliation: Fantasy Sports Interactive Ltd, Athens, Greece}
		 and {\sc Sotiris Drikos} \\[2pt]
		AUEB Sports Analytics Group, Computational and Bayesian Statistics Lab, \\
		Department of Statistics, Athens University of Economics and Business, \\
		76 Patission Street, 10434 Athens, Greece\\[6pt]
		{\rm $^{\dagger}$Corresponding author: ntzoufras@aueb.gr}\\
		%{\rm [For the Special Issue on Mathematics in Sports]}
		\vspace*{6pt}}
	%{\rm [Received on ]}\vspace*{6pt}}
	
	\maketitle
	
	\centerline{\bf Abstract}
	\begin{abstract}
		{The aim of this paper is to study and develop Bayesian models for the analysis of volleyball match outcomes as recorded by the set-difference.
			Due to the peculiarity of the outcome variable (set-difference) which takes discrete values from $-3$ to $3$, we cannot consider standard models based on the usual Poisson or binomial assumptions used for other sports such as football/soccer. Hence, the first and foremost challenge was to build models appropriate for the set-differences of each volleyball match. Here we consider two major approaches: 
			a) an ordered multinomial logistic regression model and 
			b) a model based on a truncated version of the Skellam distribution. 
			For the first model, we consider the set-difference as an ordinal response variable within the framework of multinomial logistic regression models. Concerning the second model, we adjust the Skellam distribution in order to account for the volleyball rules. 
			We fit and compare both models with the same covariate structure as in \citeasnoun{kar.ntzouf.2003}. 
			Both models are fitted, illustrated	and compared within Bayesian framework using data from both the regular season and the play-offs of the season 2016/17 of the Greek national men's volleyball league A1. 
		}
		{volleyball, Set-difference, Bayesian modelling, Multinomial logistic, Truncated Skellam.}
	\end{abstract}
	
	\pagestyle{headings}
	\markboth{}{\rm For the Special Issue on Mathematics in Sports}
	
	{\bf Acknowledgements}
	This research is financed by the Research Centre of Athens University of Economics and Business, in the framework of the project entitled ``Original Scientific Publications 2019''.

	\color{black}
	\section{Introduction and background information}
	
	Analytics and modelling of sports outcomes  have gained increased popularity over the last years due to the information and the data that are now readily available either on the web or via specialized software.
	In terms of fans' popularity, football (soccer) and basketball are the most popular sports in Europe. 
	However, volleyball, which is the sport of interest in this paper, holds also a prominent position among the rest of the team sports.
	
Volleyball belongs to the category of ``net and ball'' games. 
	The main feature of net and ball games is the absence of physical interaction between the opponent teams or players. The playing field is divided by a net which specifies the area of each team or individual player. In all net and ball games, the final outcome is recorded in many scoring stages/levels (hierarchical structure) with the final score to be reported in terms of ``sets''. 
	%The main features of this category of sports are: the set points and the absence of physical interaction. 
More specifically, the hierarchical structure of volleyball match outcomes in terms of the final score can be essentially summarized by the following procedure: 
	the team that wins the first 25 points with a margin of two points difference
	 from the opponent, wins the set.
	The team that wins three sets wins the match.
	In the case that the set-score is equal to 2--2 (two gained sets for each team), 
	then the two teams compete in a final set which is called  ``tie-break". 
	The required points-won in the ``tie-break'' is 15 in contrast to the 25 points of a usual set. 
	Hence, the team that reaches first the 15 points in the ``tie-break'' (again with a margin of two points difference) is the final winner of the match.
	These rules imply a series of important attributes that characterize volleyball. 
	First of all, no draw is allowed, which is not typical in many team sports but it is quite standard in net and ball games. 
	Second, the number of sets-won by each team are random variables that range from zero to three. 
	Finally, the set-difference determines also the number of league points earned by the two competing teams. 
	To be more specific, when a team wins with a difference of three or two sets (that is, 3--0 or 3--1) three points are attributed to the winning team and no point to the losing team. 
	In the case of 3--2 final score, then two points are attributed to  the winning team and one to the losing team. 
	Despite the popularity of volleyball in several countries, a very limited number of  statistical or mathematical approaches have been developed for modelling volleyball match outcomes. 
	We suspect that this is due to the hierarchical structure of a volleyball match and the complexity of the rules that one should take into consideration during the development of an appropriate model.

	\subsection{ Current literature on modelling volleyball  matches}

	Unlike other sports, the number of attempts for modelling volleyball match outcomes is rather limited. 
	One of the early works on the analysis of volleyball data is the article by \citeasnoun{Optimal Scoring System}. In this work, four different scoring systems designed for volleyball were compared and evaluated. 	
	Before we proceed with the review of the main research articles that focus  on modelling the final outcome of  volleyball matches, we should mention some volleyball related publications on performance analysis. 
	For example, \citeasnoun{Giatsis-Zahariadis}	
	studied and compared any possible discrepancies in the performance between two beach volleyball teams by using t-tests and discriminant analysis in order to evaluate the effect of of specific volleyball skills on the winning probability of a match. %In their analysis, they used as covariates the differences of specific volleyball skill events executed by two competing teams in each recorded beach  volleyball match. 
	Similar are the publications by  
	\citeasnoun{Florence et al}, 
	\citeasnoun{Miskin et al} and \citeasnoun{Drikos et al}. 
	In these research studies, the authors used more sophisticated methodologies and models in order to understand the effect of each skill and match characteristic on the final outcome. 
	This type of research is outside the scope of this paper and therefore we did not pursue this issue further here; see also Section \ref{Discussion} for a more detailed discussion and possible future research as well as extensions of previously introduced methodologies.  
	
	Concerning the modelling of the final volleyball outcomes, 
	one of the most notable publications is the one by  \citeasnoun{ferrante2014winning} where the winning probabilities of the set and/or the final volleyball match outcome have been estimated within a Markovian framework. In their work,  they estimated the winning set probabilities for both rally scoring systems (the current one and the older one with 25 and 15 points-won per set, respectively) as well as 
	the expected duration of a set (measured by the number of played rallies).

	The logistic regression model is a simpler but effective alternative approach for modelling a binary final match outcome (in the form of win/loss) between two competing teams in sports with no draws (as in volleyball). 
	%A simpler alternative approach for modelling the final match outcome (win/loss) between two competing teams in sports with no draws (as in volleyball) is to use the logistic regression model. 
	Following this approach, \citeasnoun{akarccecsme2017possible} presented an analysis of the final  volleyball match outcome (win/loss)    by using as covariates the players' efficiency for each role/position as described by \citeasnoun{Afonso et al}; as efficiency they considered the difference between  scored and conceded points.  
	\citeasnoun{Zhang} also followed the same approach using a collection of in-game statistics as covariates. 
	He further implemented a simple regression model for modelling the point difference leading to a model which explains $92\%$ of the response with only two in-game characteristics; 
	see also in \citeasnoun{Magel_etal}.
	Although such analysis can provide useful insights about the volleyball game, 
	both of these modelling approaches have disadvantages when compared to sophisticated models. 
	The use of  binomial regression models for analysing volleyball outcomes  
	does not allow us to predict the final league ranking  points of each team since the exact set-difference remains unknown, while the simple regression model on the point difference overlooks the basic characteristics of the game itself and it cannot be used for accurate predictions and inference since the structural model assumptions (such as the normality of the response) are violated. 
	
	From the machine learning perspective, \citeasnoun{AI-NN} implemented an Artificial Neural Network (ANN)  approach to predict the final league ranking based on several data such as the number of wins and defeats of each team. 
Although the method demonstrated impressive predictive performance, this approach does not offer any useful information about the game itself due to  the   lack of interpretability of this approach.

	A popular choice for modelling data from team sports with a discrete score measured in goals such as football (soccer) or water polo are  models based on Poisson regression and their extensions \citep{lee1997,kar.ntzouf.2003}. 
	Clearly, such models are not appropriate for modelling the sets in volleyball match which take values in $\{ 0, 1, 2, 3\}$. 
	Nevertheless, a Poisson model with additional hierarchical structure  was used  by \citeasnoun{Gabrio} for modelling the points of the two competing teams in a volleyball match. 
	He proposed a three level Bayesian hierarchical formulation which modelled simultaneously the points (using a simple Poisson component), the probability of winning a set (using a binomial logistic regression component) and the probability of winning a match. 
	%The last two hierarchical levels, used the predicted points from the first level as a predictor.  
	The work of \citeasnoun{Gabrio} is a more sophisticated treatment of volleyball   match   outcomes (in comparison with simpler models used in the past) by considering specific features and characteristics of the game. 
	Again a shortcoming of this work is that such model formulation does not take into consideration that the points of each team for each set are discrete bounded correlated random variables and this cannot be described by the use of a double (independent) Poisson model.

	A similar but more elaborate approach for modelling jointly the set and the points was recently introduced by \citeasnoun{egidi-ntzoufras}. In this work, a top-down Bayesian modelling approach of both sets and points by taking into account simultaneously other game characteristics such as the extra points needed to be played because of the two points of margin rule of win. 
	Their proposed model is a two-level hierarchical model, but the model component for the points is a sophisticated truncated negative binomial model taking into consideration all the volleyball specific features and point rules. 
\subsection{Our modelling strategy}
	Here we take a different research path and we focus on modelling the  set-differences in volleyball. 
	We focus on two different modelling approaches: (a) the ordered multinomial logistic regression model and (b) a model based on a truncated version of the Skellam distribution.
	Both of these models take into consideration the main rules and characteristics of volleyball 
	as described at first paragraphs of this section which could not be handled by models based either on the binomial or the Poisson distribution adopted by \citeasnoun{lee1997,kar.ntzouf.2003} and partially by \citeasnoun{Gabrio} for the points of each set.  
		
		The first model is an ordered multinomial logistic regression model. 
		This is a standard extension of the multinomial logistic regression model which accounts for the ordinality of the six possible outcomes $\{ -3,-2,-1,\,1,\,2,\,3 \}$. 
	
	The second model is a modification of the  Skellam (or Poisson difference) model used by \cite{kar.ntzouf.2008} to model the difference of goals directly in football (soccer). 
	Since, the Skellam distribution is defined in the whole set of integer numbers, it cannot be used for modelling  sufficiently the set-differences which take values in $\{-3,-2,-1,1,2,3\}$. 
	Hence, we introduce a truncated version of the Skellam distribution, that accounts for the constrained range of the response variable of the set-difference. 
	Moreover, modelling the difference of sets in volleyball has an important benefit in comparison with football or other invasion team sports with only one level of scoring points (i.e. goals). 
	In volleyball, if the set-difference is known then the final set result of the corresponding match is also available since differences of $\{-3,-2,-1,1,2,3\}$ correspond to scores \{0--3,\,1--3,\,2--3,\,3--2,\,3--1,\,3--0\}, respectively. 
	So essentially, no information is lost by modelling the set-difference in contrast to football where the exact score can not be recovered by modelling only the difference. 
	Hence, the idea here is to use an appropriate adaptation of the Skellam distribution in order to build a tailor made model/distribution for volleyball set-differences.
	
	Both modelling approaches will be used for prediction and interpretation of the game itself. 
	The aim of this paper is to validate and examine the appropriateness of these two models for set-differences. 
	Both of the models are implemented within the Bayesian framework. 
	We use Markov chain Monte Carlo (MCMC) algorithm  \citep[see for example in][]{MCMC}   to estimate posterior densities and also reproduce the final league standings as well as  evaluate both the goodness of fit and the prediction accuracy of the proposed models.

\subsection{Outline of the paper}
	
	The rest of this paper is organized as follows: Section 2 introduces the proposed models and describes in detail some useful preliminary theoretical properties which provide intuition and better understanding of  model parameters. 
	In Section 3, we  implement the proposed methods and models to data from the Greek A1 professional men's league of the regular season 2016/17. We provide full interpretation of  model parameters. Focus is given on the estimated attacking and defensive abilities. 
	In Section 4, we use the posterior predictive distribution, obtained within the implemented MCMC runs, 
	in order to re-generate the league and evaluate the fit of the model. 
	Moreover, by using again the posterior predictive distribution, we evaluate the predictive power of the proposed models under two scenarios: 
	(i) in a mid-season (split-half) prediction scenario where the matches of the first half of the season are used for estimation/learning while the rest of them are used as a test/validation dataset, and 
	(ii) in a play-offs prediction scenario, where the data of the regular season along with the data of previous play-off phases  (if they exist)   are used to estimate the final results of each match of the play-off rounds. 	
	Finally, Section  5   discusses the findings of this paper and provides recommendations for further research.

	\section{Bayesian modelling for the set-difference}
	
	\subsection{Dataset and league specifics}
	The  methodologies  we propose in this article have been implemented on data from the  Greek volleyball league (also called A1 Ethniki) for season 2016/17	which is the highest professional volleyball league in Greece. 
	It is run by the Hellenic Volleyball Federation and it is considered one of the top national leagues in European volleyball, as its clubs have had significant success in European competitions.
	The data are subset of a larger dataset collected via a volleyball specific software where scouts were registering every touch of the ball during the match. 
	For each  set in the entire league,  we record the data for both competing teams of the A1 men's regular season 2016/17. 
	All data have been collected by the third author of this manuscript 
	% **** Double blinded info
	%(Dr. Sotiris Drikos 
	(former manager of the Greek National volleyball team and expert on performance analysis).
	
	The regular season data consist of $n=132$ matches with 494 sets (in total). 
	Twelve ($p=12$) teams were involved in this league which they compete against each other every week (match day). During the regular season, each club competes with the rest of the teams twice in a double round-robin system, once at their home stadium and once at their opponents stadium. Hence, the regular season league has 22 match days and in every match day there are six matches. 
	The regular season is followed by the matches of the play-off phases between the top eight teams of the regular season and by the matches of the play-out phases between the bottom three teams finishing at positions 9--11. 
	The final league ranking is announced after the completion of play-offs and play-out phases. 
	According to this final ranking, the first team wins the championship title while the champion along with the second finalist team qualify to play in the Champions league European competition. 
	The third and fourth teams qualify to play in the CEV Cup and the Challenge Cup, respectively, which are the second and third-tier level competitions for men's volleyball clubs of Europe. 
	The last team (position 12) at the end of the regular season relegates directly to the second division.

	\subsection{The ordered multinomial logistic model}
	
 First, we consider the standard ordered multinomial formulation for modelling the response $Y$ which  here is defined as the difference between the sets of the two competing teams in each match \color{black}. 
	This set-difference will be treated as our  ordinal response variable with $K=6$ levels which are all the potential match \color{black} outcomes. 
	Hence, for an observed set-difference equal to $k$, the response ($Y$) is defined as 
	\begin{equation}\label{def_res_mul}
	y^{(k)}= k-3- \mathcal{I}( k \le 3 )
	\end{equation}
	for $k= 1, \ldots, 6$;  where $\mathcal{I}( k \le 3 )$ takes the value of one when $k \le 3$ and zero otherwise. 
	Hence, $Y \in \{-3,-2,-1,1,2,3\}$, while $y^{(1)}$ and $y^{(6)}$ refer to the set-difference of $-3$ and $3$ (i.e. final match \color{black} scores $0-3$ and $3-0$ sets), respectively. 
	The multinomial distribution is specified by
	\begin{equation}\label{mult-volleyball}
	\Phi(Y_{i}) \sim Multinomial(n_{i}=1, \boldsymbol{\pi}_i)\\
	\end{equation} 
	where $\Phi(Y)=Y+3+\mathcal{I}( Y \le 0 )$, $ \boldsymbol{\pi}_i=(\pi_{i1},\ldots, \pi_{iK})$ is the vector probability of $K=6$ possible set-differences for $i= 1,\ldots, n=132$ matches \color{black}  and $\sum_{k=1}^{K}\pi_{ik}=1$. In other words, the $\pi_{ik}=P(Y_{i}=y^{(k)}$) is the probability of set-difference being equal to $y^{(k)}$ in match \color{black} $i$.

	In our application, the linear predictor of the ordered-multinomial model is specified by
	\begin{equation}\label{ordered-volleyball}
	\begin{split}
	\log\frac{\gamma_{ik}}{1-\gamma_{ik}} &=c_{k}-(A_{ht_{i}}-A_{at_{i}})
	\end{split}
	\end{equation} 
	where $\gamma_{ik}=P(Y_{i}\leq y^{(k)})=\sum_{j=1}^k \pi_{ij}$ is the probability that the response outcome falls in category $k$ or in a lower category in match $i$, for $i=1,\ldots, 132$ and $k= 1,\ldots, K-1$; $c_{k}$ are constant parameters for $k=1, \ldots, K-1$ ($c_{k} < c_{k+1}$) and $A_{ht_{i}}$ as well as $A_{at_{i}}$ are the ``net" general abilities of the home and the away team  in match $i$ (denoted by $ht_i$ and $at_i$, respectively). 
	In the linear predictor \eqref{ordered-volleyball}, we consider the difference between the general abilities of the home and the away team since this will mainly determine the probability of winning and the final set-difference in a match.
	The negative sign in front of the abilities difference in (\ref{ordered-volleyball}) is adopted in order to facilitate the interpretation of model parameters. Hence, by this formulation, a large difference in the abilities of two competing teams will result in an increase to the probabilities of larger differences in terms of sets.  
	This model is called proportional odds model \citep{Agresti2013}. For the general ability parameters, we adopt a sum to zero constraint    
	$$\sum_{j=1}^{p}A_j =0$$ 
	in order to express the ability of each team as a deviation from the performance or ability of an average team; where $p$ denotes the number of teams in the league (in our case $p=12$). 
	The general ability of the omitted team is calculated by 
	\begin{equation}\label{omit_team}
	A_{1}= - \sum_{j=2}^{p}A_{j}.
	\end{equation}
	As far as the specification of prior distributions is concerned, low informative priors for all parameters such as the normal distribution with mean equal to zero and large variance (e.g. $10^{4}$) are adopted since here we have assumed that no historical information was available.  
	
	%\subsection{A Skellam's variation model}
	\subsection{The Zero-Deflated and Truncated Skellam model}
	
	\subsubsection{Model formulation.} ~\\ 
	\label{SectionZDTS}
	Here we use the Skellam (or Poisson difference) distribution as the basis for modelling the set-differences directly; see, for example, \citeasnoun{kar.ntzouf.2008}. For this reason, it is necessary to mention briefly some details and properties of this distribution before proceeding with the presentation of our model based on a variation of the Skellam distribution. 
	
	\citeasnoun{irwin1937frequency} introduced the distribution of the difference of two independent Poisson random variables for the case of equal means. After a decade, \citeasnoun{skellam1946frequency} moved one step forward by specifying the case for unequal means. 
	More recently,  \citeasnoun{karlis2006bayesian} introduced the distribution of the difference between two 
	correlated random variables that follow the bivariate Poisson distribution. 
	More specifically, assuming that $X=W_{1}+W_{3}$ and $Y=W_{2}+W_{3}$ with $W_{1}\sim Poisson(\lambda_{1})$, $W_{2}\sim Poisson(\lambda_{2})$ and the $W_{3}$ follows any discrete distribution with parameter $\lambda_{3}$, they resort to the apparent result: $Z=X-Y=W_{1}-W_{2} \sim Sk(\lambda_{1},\lambda_{2})$, which does not depend anymore on $W_{3}$ and consequently the correlation is eliminated. 
	In essence, the joint distribution of $(X,Y)$ is a bivariate distribution incorporating the correlation through the variable $W_{3}$. In the case that $W_{3}$ is also a Poisson random variable, both marginals $X$ and $Y$ follow Poisson distribution.  The probability function of $Z$ is given by: 
	\begin{equation}\label{skellamdistr}
	f_{Sk}(z|\lambda_{1},\lambda_{2})=P(Z=z|\lambda_{1},\lambda_{2})=e^{-(\lambda_{1}+\lambda_{2})}\left(\frac{\lambda_{1}}{\lambda_{2}}\right)^{\frac{z}{2}}I_{|z|}\left(2\sqrt{\lambda_{1}\lambda_{2}}\right)
	\end{equation}
	with $z \in \mathbb{Z} $, $\lambda_{1},\lambda_{2}>$0 and $I_{r}(x)$ is the modified Bessel function of order $x$ of the form 
	\begin{equation*}
	I_{r}(x)=\left(\frac{x}{2}\right)^{r}\sum_{k=0}^{\infty} \frac{\left(\frac{x^{2}}{4}\right)^{k}}{k!\Gamma(r+k+1)}; 
	\end{equation*}
	see \citeasnoun{abramowitz1965handbook} for more details. 
	The support of this random variable is the set of integer numbers and can be used for modelling differences between discrete counts. The main difference between the two cases of independent and the dependent Poisson variates  is the interpretation of the parameters $\lambda_1$ and $\lambda_2$. 
	The expected value and the variance of $Z$ are given by
	\begin{equation} \label{Exp_Sk}
	\begin{aligned}
	E(Z_{Sk})&=\lambda_{1}-\lambda_{2}\\
	Var(Z_{Sk})&= \lambda_{1}+\lambda_{2}.
	\end{aligned}
	\end{equation}
	% There are plenty of other useful properties for this distribution which are not to be mentioned in this work because they are not of our primary importance (for more details, see \citeasnoun{karlis2006bayesian}).
	
	The Skellam distribution is useful for modelling sport outcomes, since, by this way, we can eliminate any linear  correlation between the score outcomes of the two competing teams. 
	When modelling the set-differences of volleyball matches, essentially we model the full set-score. 
	This is in contrast to other team sports such as football or basketball where using the goal or point difference is not equivalent to modelling the full time match score. This is because portion of the final score information is lost  by using only the differences. 
	However, the use of the Skellam distribution for modelling set-differences,   
	in a similar manner as \citeasnoun{kar.ntzouf.2008} used this model for goal differences in football,
	is not appropriate.  
	This is due to the restrictions imposed on the set-difference by the scoring rules of the volleyball game. 
	
			%*** 13/9/2020 *** stopped here **** 
First, the random variable of our response (the set-difference) cannot be equal to zero (draw). Second, the set-difference cannot be greater than three or smaller than minus three. The first case refers to the non-existence of ties (draws) in a volleyball match while the latter case refers to the fact that one of either home or away team wins with maximum three sets margin, respectively.  In other words, the response $Z=X-Y$ of set-difference (with $X$ and $Y$ being the sets scored by the home and the away team, respectively) is strictly defined to $\{-3,-2,-1,1,2,3\}$. The negative values of this support correspond to the win of the away team and the positive correspond to the win of the home team. For this reason, a new modified model based on the Skellam distribution is proposed in order to take into consideration the above mentioned constraints. To fit this new version of Skellam model, we have  to define the corresponding distribution of this model. Hence, we define the \textbf{ zero-deflated and truncated Skellam distribution (ZDTS)} as the one with probability mass function
	\begin{eqnarray} \label{ZDTS}
	\begin{aligned}
	f_{ZDTS}(z|\lambda_{1},\lambda_{2}) & =& P(Z=z|Z \in \{-3,-2,-1,1,2,3\}) \\
	&  =&\frac{f_{Sk}(z|\lambda_{1}, \lambda_{2})}
	{  \sum \limits_{ x \in \{1,2,3\}}\ \Big\{ f_{Sk}(-x|\lambda_{1}, \lambda_{2}) + f_{Sk}(x|\lambda_{1},   \lambda_{2}) \Big\}}   
	\end{aligned}
	\end{eqnarray}
	for $z \in \{-3,-2,-1,1,2,3\}$. 
	As you may observe in \eqref{ZDTS}, the truncation of the Skellam distribution is in both minus four (lower truncation) and four (upper truncation).
	
	Hence, the zero-deflated and truncated version of the Skellam distribution is used for the modelling of set-differences \begin{equation}\label{ZDTS_distr}
	Y_{i}\sim ZDTS(\lambda_{i1}, \lambda_{i2})
	\end{equation}
	where $Y_{i}$ is the set-difference in match \color{black} $i$, for $i= 1, \ldots, 132$ regular season matches.
	%%%%%%%%%%%%%%%%%%%%%%%%%%%%%
	
	For this model we have used the linear predictor formulation 
	\begin{equation} \label{ZDTSLIN}
	\begin{aligned}
	\log({\lambda_{i1}})  &= \mu+home+att_{ht_{i}}+def_{at_{i}}\\
	\log({\lambda_{i2}}) &= \mu+att_{at_{i}}+def_{ht_{i}}
	\end{aligned}
	\end{equation}
	where $\mu$ is a constant parameter, $home$ is the common home effect, $ att_{ht_{i}}$ as well as $att_{at_{i}}$ are the ``net" attacking abilities of the home and the away team in match \color{black} $i$.  In the same way,   
	$def_{ht_{i}}$ and  $def_{at_{i}}$  are used to capture the ``net" defensive abilities of the home and the away team  in match $i$. For both ability parameters, sum-to-zero constraints (as in the ordered-multinomial model) are used  
	\begin{equation} \label{eq2}
	\sum_{j=1}^{p}att_j =0 \mbox{~and~} \sum_{j=1}^{p}def_j =0
	\end{equation} 
	where $j= 1, \ldots, p$ denotes the team $j$ participating in the league.
	Both abilities of the omitted team are calculated as in \eqref{omit_team}.

	In the above formulation we have used the standard expression connecting the team ability parameters directly with  $\lambda_1$ and $\lambda_2$ rather than the mean of ZDTS distribution; 
	see \citeasnoun{kar.ntzouf.2008} for a similar treatment when using the original Skellam distribution. 
	This was mainly due to the complexity of the expected value of $Z$ given by 
	%\eqref{mean_zdts}, 
	%Hence, the expected value for the random variable Z of the zero-deflated and truncated Skellam distribution is specified by
	\begin{equation}\label{mean_zdts}
	E(Z_{ZDTS}) =\left(\lambda_{1}-\lambda_{2}\right)\frac{I_{1}+\frac{2I_{2}\left(\lambda_{1}+\lambda_{2}\right)}{\left(\lambda_{1}\lambda_{2}\right)^{\frac{1}{2}}}+\frac{3I_{3}\left(\lambda_{1}^{2}+\lambda_{1}\lambda_{2}+\lambda_{2}^{2}\right)}{\lambda_{1}\lambda_{2}}}
	{I_{1}\left(\lambda_{1}+\lambda_{2}\right)+\frac{I_{2}\left(\lambda_{1}^{2}+\lambda_{2}^{2}\right)}{\left(\lambda_{1}\lambda_{2}\right)^{\frac{1}{2}}}+\frac{I_{3}\left(\lambda_{1}^{3}+\lambda_{2}^{3}\right)}{\lambda_{1}\lambda_{2}}}
	\mbox{~with~} 
	I_{\kappa}=I_{\kappa}(2\sqrt{\lambda_{1}\lambda_{2}}) \mbox{~for~} \kappa= 1, 2, 3,   
	\end{equation} 
	which makes the direct modelling of $E(Z_{ZDTS})$ extremely  difficult. 
	Nevertheless, the above expression provides us a useful insight about the expected value of the ZDTS  distribution which is proportional to the mean of the original Skellam distribution (given by $\lambda_1-\lambda_2$) which is now multiplied by a correction factor which keeps the expectation  $E(Z_{ZTDS})$   within the range of acceptable values (i.e. between $-3$ and $3$).

\begin{figure}[h]
		\centering
		\includegraphics[width=0.8\textwidth,height=1\textheight,keepaspectratio]{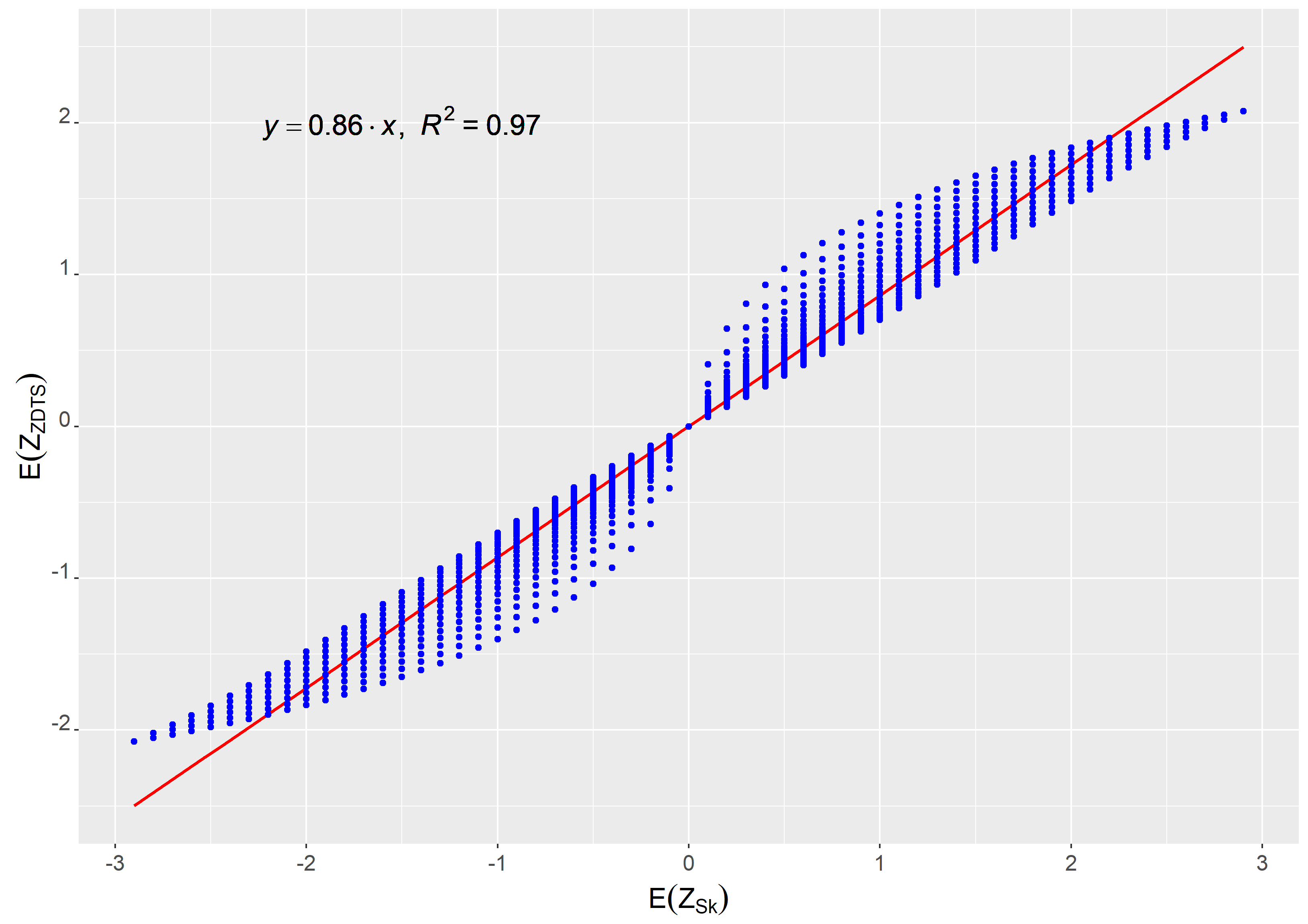}
		\caption{The linear relationship between the expected values $E(Z_{ZDTS})$ and $E(Z_{Sk})$; These expectations are denoted as $y$ and $x$, respectively, in the equation at the top-left of the figure. The 
		high $R^2=0.97$	implies that the simplified ZDTS interpretation can be used satisfactorily.}
		\label{exp_zdts_3}
	\end{figure}

	According to Figure \ref{exp_zdts_3}, the expected values of $E(Z_{Sk})$ and $E(Z_{ZTDS})$ are quite close for a variety of value combinations of  $(\lambda_1, \lambda_2) \in [-3,3]^2$. 
	This implies that the interpretation based on the simplified model formulation  of the ZDTS model    will be similar to the corresponding interpretation of the parameters for the original Skellam distribution. 
	
	\subsubsection{Prior specification.}~\\  
	\label{ZDTS_prior}
	The prior specification for the parameters of this model is not a straightforward task since we have observed numerical problems related with the scale of $\lambda_1$ and $\lambda_2$. 
	This is because \color{black} several combinations of values for $\lambda_1$ and $\lambda_2$ will result to unrealistic set-differences.  
	For this reason, we have calibrated the prior distributions of  $\lambda_1$ and $\lambda_2$ in order to be roughly in agreement with the model assumption: $-3 \leq E(Z_{ZDTS}) \leq 3$. 
	Using the latent variable interpretation of the Skellam distribution, and in order to restrict the scaling of $\lambda_1$ and $\lambda_2$ we have decided to use the assumptions of $0\leq \lambda_{1}\leq3$ and $0\leq \lambda_{2}\leq3$.
	\newcommand\myeq{\stackrel{\mathclap{\normalfont\mbox{($\mu=0$)}}}{\Rightarrow}}
	
	\begin{equation*}
	\openup\jot % make lines a little more far apart
	\begin{aligned}[t]
	&  0\leq \lambda_{2}\le 3\\
	\Rightarrow  & 0 \leq \exp(\mu)\leq 3\\
	\Rightarrow  & -\infty \leq \mu \leq 1.10
	\end{aligned}
	\qquad\qquad % adjust to suit
	\begin{aligned}[t]
	&0\leq \lambda_{1} \leq 3\\
	\Rightarrow &0 \leq \exp\left(\mu + home\right) \leq 3\\
	\Rightarrow & -\infty \leq \mu + home \leq 1.10\\
	\end{aligned}
	\end{equation*} 
	Therefore, the value of $1.10$ can act as a rough upper bound for  both the constant parameter $\mu$ and the home effect. 
	Following this logic, we have used the three-sigma rule \citep{pukelsheim1994three} for $\mu$ and $home$ 
	in order to specify their prior standard deviation. 
	Moreover, we assume prior variance equal to one for the ability parameters. 
	Although smaller prior variances are also plausible, we have adopted this choice to ensure that our analyses will not be overly influenced by the choices of priors for $\lambda_1$ and $\lambda_2$. 
	Hence, the prior distributions we have finally used in our approach are the following:   
	\begin{equation*}\label{priors ZDTS}	   
	\openup\jot % make lines a little more far apart
	\begin{aligned}[t]
	\mu &\sim N(0,0.37^{2}),  \\ home &\sim N(0,0.37^{2}), 
	\end{aligned}
	\qquad\qquad % adjust to suit
	\begin{aligned}[t]
	att&\sim N(0,1^{2}), \\           def&\sim N(0,1^{2}). 
	\end{aligned}
	\end{equation*} 
	Sensitivity analysis  have shown that posterior results are quite robust over different choices of prior variance values, indicating that our suggested prior is indeed a low information prior; 
	see Section \ref{prior_sens} for more details.

\subsection{Computational implementation and posterior estimation details}
The posterior results for all models fitted in this paper have been obtained by running a Hamiltonian Monte Carlo (HMC) algorithm  \citep{HMC} using  the rstan package \citep{Rstan} in {\sf R} \citep{R Core Team} which is an interface to a probabilistic Bayesian programming language called \Stan \citep{Carpenter et al,Stan}. 
  More specifically, \Stan uses
a generalization of the familiar Metropolis algorithm, the No-U-Turn Sampler
(NUTS), which is an adaptive invariant of HMC, in order to obtain a simulated sample from the target distribution; for more details about NUTS see \cite{NUTS}.

	\section{Preliminary analysis for the Greek volleyball league 2016/17}
	
	Before proceeding to the main results of our proposed methodology, we first present preliminary results concerning the structure of the linear predictor in the ZDTS model (Section \ref{structureZDTS}), 
	the prior sensitivity analysis (Section \ref{prior_sens}) and 
	the MCMC convergence diagnostics (Section \ref{convergence_diag}) 
	of our proposed models implemented on the Greek volleyball league data of season 2016/17. 
	These analyses were required in order to ensure 
	(a) the use of the appropriate structure of the implemented ZDTS model (and the identifiability of the proposed model), 
	(b)	the objectivity and the stability of the results concerning the prior choice and 
	(c) the generation of values from the correct posterior distribution.

	\subsection{Comparing models with different structure of the linear predictor of the ZDTS model}  
	\label{structureZDTS}

	In this section we provide details and comparisons of models with different team ability structure. 
	The team abilities are estimated via parameters introduced in the linear predictors of   $\lambda_1$ and $\lambda_2$ of the ZDTS model. 
	The model introduced in Section \ref{SectionZDTS} is considered as the basic/reference model (referred in the following as model 1). 
	It assumes different attacking and defending team abilities  and the formula of the linear predictor is given by \eqref{ZDTSLIN}. 
	This model is compared with three alternative models assuming common team abilities. 
The general linear predictor of these models can be summarized as 
\begin{equation} \label{ZDTS_gen}
\log(\lambda_{i1}) = \mu+home+ \ell_{i1} \mbox{~~and~~} 
\log(\lambda_{i2}) = \mu+ \ell_{i2}~. 
\end{equation}
Parameters $\ell_{i \kappa}$ represent the team ability structure introduced in the linear predictors of parameters $\lambda_{i \kappa}$ (for $\kappa=1,2$). 
Model 1 can be  summarized in the form  of \eqref{ZDTS_gen} with 
$\ell_{i1}=att_{ht_i}+def_{at_i}$ and  $\ell_{i2}=att_{at_i}+def_{ht_i}$. 
Here we consider three additional models with $\ell_{ij}$  given by 
\begin{equation*} \label{ZDTS2}
\begin{aligned}
\mbox{Model 2:} &  & \ell_{i1} &= ability_{ht_{i}} &\ell_{i2} &= ability_{at_{i}} \\ 
\mbox{Model 3:} &  & \ell_{i1} &= ability_{ht_{i}}-ability_{at_{i}} &\ell_{i2} &= 0 \\ 
\mbox{Model 4:} &  & \ell_{i1} &= 0 &\ell_{i2} &= ability_{at_{i}}-ability_{ht_{i}} ~. 
\end{aligned}
\end{equation*}

For Models 2--4 we have used similar prior distributions as the ones in Model 1 proposed in Section \ref{ZDTS_prior}. 
The Bayesian predictive measures of the Watanabe-Akaike information criterion  \citep[WAIC]{WAIC} and the leave-one-out information criterion \citep[LOOIC]{LOO--WAIC} are presented in Table \ref{1Comparison_ZDTS} showing that the model with different attacking and defensive abilities on $\lambda$s (Model 1) is preferable to the three alternatives considered here. 
We have also run Models 2--4 with more dispersed priors for  sensitivity reasons. Results were identical (with small changes in Model 2) suggesting that Model 1 is again better than its competitors according to both WAIC and LOOIC (details are omitted for brevity).

\begin{table}[h]
	\begin{center} 
	\begin{tabular}{ c ll c c }
		\toprule
		Model & $\ell_{i1}$ & $\ell_{i2}$ &    WAIC & LOOIC \\ 
		\midrule
		(1)&  $att_{ht_{i}}+def_{at_{i}}$ & $att_{at_{i}}+def_{ht_{i}}$           &\textbf{379,2}& \textbf{379,9}\\ 
		(2)& $ ability_{ht_{i}}$ & $ability_{at_{i}}$                                    & 385,4 & 385,8 \\
		(3)& $ability_{ht_{i}}-  ability_{at_{i}}$ & $0$                                 & 384,0 & 384,3 \\
		(4)& $0$                                   & $ability_{at_{i}}-ability_{ht_{i}}$ & 383,6 & 384,1 \\
		\bottomrule
		\multicolumn{5}{p{10cm}}{\footnotesize \it The linear predictors for all models can be summarized as 
			$\log(\lambda_{i1})= \mu + home + \ell_{i1}$ and
			$\log(\lambda_{i1})= \mu + \ell_{i2}$ with $\ell_{ij}$  provided in columns 2--3; 
			the subscripts $ht_i$ and $at_i$ refer to the home and away teams playing in match $i$.}
	\end{tabular}
	\caption{Comparison between four candidate fitted models based on the ZDTS distribution by using WAIC and LOOIC. All priors are specified as: mu, home$\sim$N(0,$0.37^{2}$), att, def, ability$\sim$N(0,$1^{2}$); Burnin $B=1000$, Iterations Kept $T=5000$,}
	\label{1Comparison_ZDTS}
	\end{center} 
\end{table}

	\subsection{Prior sensitivity analysis}  
	\label{prior_sens}
	
	Section \ref{ZDTS_prior} describes in detail the specification of the prior distributions of the model parameters in order to avoid 
	(a) possible non-identifiability problems and (b) 
	post-warmup divergence errors appearing in {\sf RStan} due to large differences in $\lambda_{i1}$ and  $\lambda_{i2}$. 
	
	Here, following a comment by a referee, we have implemented our basic ZDTS model with linear predictor \eqref{ZDTSLIN} by using different prior variances. 
	We have conducted three different sensitivity experiments 
	for (a) the constant parameter $\mu$ (b) the home effect ($home$)
	and (c) the team ability parameters (both $att$ and $def$ simultaneously). 
	For each implementation, we have adopted the prior setup proposed in Section \ref{ZDTS_prior} as a baseline/starting point. 
	Then, we multiply the proposed prior standard deviation of the parameter(s) examined in each sensitivity experiment by $c \in \{2, 5, 10, 20\}$. For each different prior combination, we have estimated the posterior model deviance measure which can be considered as an overall measure of fit across the same model for different prior setups. 
	
	Table \ref{sensitivity_parameters_att_def_mu} presents the posterior mean deviances along with posterior standard deviations for different prior sensitivity exercises. 
	For all parameters, it is clear that the deviance measure is robust, keeping its posterior mean value stable even if we multiply the proposed prior standard deviation by the value of $20$. 
	Nevertheless, for parameter $\mu$ and the team abilities, 
	\Stan reported post-warmup divergence warnings which were increasing  along with $c$  indicating that results of some specific iterations may not be reliable (in practice large parameter values cause overflows for specific iterations). 
	
	To conclude with, our prior sensitivity experiments have not indicated any sensitivity issues suggesting that the selected priors are generally weakly informative leaving the data to dominate the posterior. Moreover, the proposed prior values of Section \ref{ZDTS_prior} avoid post-warmup divergence problems reported by \Stan. 
	Finally, the overall fit and the predictive performance of the implemented ZTDS model (presented in Section \ref{main_results} which follows) also re-assures us that the proposed model setup is in the correct direction. 

	\begin{table}[ht]
		\centering
		\small 
		\begin{tabular}{l rr@{~}l @{~~} r@{~}l @{~~} r@{~}l @{~~} r@{~}l}
						\hline
			Parameter    & \multicolumn{1}{l}{$c=1$} & \multicolumn{2}{l}{$c=2$}& \multicolumn{2}{l}{$c=5$}& \multicolumn{2}{l}{$c=10$}& \multicolumn{2}{l}{$c=20$}\\  
			\hline
			$home$       & 358.7(6.5) & 358.6~(6.6)&& 359.5~(6.7)  && 359.2~(6.4) && 358.9~(6.7)& \\
			\hline
		
			$\mu$        & 358.7(6.5) & 356.8~(6.0)&& 356.9$^*$(6.2)& [0.9\%] & 356.8$^*$(6.0)& [~0.5\%] & 357.9$^*$(5.9) &[13.7\%] \\  
			\hline

			$att~\&~def$ & 358.7(6.5) &  357.4$^*$(6.5)&[0.2\%]& 357.9$^*$(7.0)&  [3.3\%] &  357.7$^*$(7.1) &[10.6\%]& 358.6$^*$(7.3) &[20.6\%]\\
			\hline
			\multicolumn{10}{p{14cm}}{ \footnotesize \it $^*$Within brackets: \% of iterations with post-warmup divergences reported in \Stan (results may be biased).   }\\
			\multicolumn{10}{p{14cm}}{ \footnotesize \it $c$: prior standard deviation multiplicator; $c=1$ corresponds to the prior standard deviation recommended in Section \ref{ZDTS_prior}. }
		\end{tabular}
		\normalsize 
		\caption{Posterior means (standard deviations) of the model deviance measure  over different prior variances (with all the remaining ones set at the values proposed in Section \ref{ZDTS_prior}); Burn-in $B=1000$, Iterations kept $T=1000$.}
		\label{sensitivity_parameters_att_def_mu}
	\end{table}

	\subsection{Convergence of MCMC algorithms}   
	\label{convergence_diag}
	
	The convergence of the MCMC algorithm for the ordered multinomial model 
	(Eq. \ref{mult-volleyball}--\ref{ordered-volleyball}) and the ZDTS model 
	(Eq. \ref{ZDTS_distr}--\ref{ZDTSLIN}) has been examined using the  \cite{Raftery_Lewis_1992} 
	diagnostic and graphically by using trace plots, ergodic mean plots  and auto-correlation plots. 
	All relevant MCMC output analysis is provided in the electronic appendix of the paper (see Appendix B) for MCMC runs of three chains of $T=6000$ iterations kept. For the ordered multinomial model, each chain was run for 16000 iterations for which a thinning equal to 2 was applied, and then a burn-in of 
	$B=2000$ was eliminated (from the thinned chain) resulting to the final sample of 6000 iterations. 
	In total $3\times 6000 = 18000$ iterations were kept.
	Equivalently, for the ZDTS model we have run three chains with length of $6000$ iterations (18000 in total) obtained in a similar manner but by using a thin equal to 5 (out of total of 40000 iterations). 
	This output was used in the main analysis of Section \ref{main_results} which follows. 
	Although results using a single chain of $B=2000$ and $T=6000$ with no thinning also indicated convergence, we have finally selected 3 chains and thinning of 2 and 5, respectively, in order to eliminate any auto-correlation and obtain more accurate and stable posterior estimates.

	\section{Main results for the Greek volleyball league 2016/17}	 
	\label{main_results}
	\subsection{Posterior analysis and interpretation}
	
	In this section  we focus on the interpretation of the estimated parameters of both models under consideration. 	
	We first present the results of the ordered multinomial logistic model followed by the results of the ZDTS model. We present summaries and figures based on the posterior values of the parameters of interest. 
	Tables with summaries of the posterior distributions of all model parameters are provided at Appendix C.1. 

	The constant parameters $c_1, \dots, c_{5}$ of the ordered-multinomial model (Eq. \ref{ordered-volleyball}) express the log cumulative odds ratio of each set-difference threshold to the remaining set-differences when two teams of equal strength are competing each other.  
	Posterior summaries of these parameters are presented at Table C.1 (Appendix C). 
	In order to understand their effect  and make the values comparable to the corresponding values induced by the ZDTS model, Table \ref{probab-ordered} presents the expected relative frequencies (in $\%$) for each value of the set-differences for a match between two teams of equal strength for both fitted models.
	According to this table, we observe that 
 
	\begin{enumerate}
		\item The home team has greater probability to win than the away team. \\
		This is induced by  constant parameters $c_k$. The case of no home effect is induced when $c_3=0$ and
		$|c_k|=|c_{6-k}|$ for $k=1,2$. 
		Here, clearly the probability of winning for the home team ( $P(Z\ge 1) = 0.608$ ) is greater than the corresponding one for the away team when the competing teams are of the same strength while $\pi_k < \pi_{7-k}$ for all $k=1,2,3$ since 
		$(0.058, 0.168, 0.165) \preccurlyeq (0.124, 0.196, 0.287)$.  
				
		\item The most frequent score is $3-1$. \\
		The difference of two sets in favour of the home team (i.e. 3--1) is the most frequent score with relative frequency equal to 29\% followed by scores 3--2,  1--3 and 2--3   with relative frequencies ranging from $15\%$ to $20\%$. 
		The probability of observing a difference of three sets is much lower ($\sim12\%$ and $\sim 5.5\%$ for home and away teams, respectively) which is reasonable for a match between teams of equal strength.  
	\end{enumerate}
	
	\begin{table}[h!]
		\centering
		\begin{tabular}{lcccccc}
			\hline
			& \multicolumn{6}{c}{\textbf{Set-differences}} \\ 
			\cline{2-7}
			Model           & -3 & -2 & -1 & 1 & 2 & 3 \\  \hline
			Ordered M.  & 5.84 (1.87) & 16.83 (3.87) & 16.52 (4.40)& 19.65 (4.82) & 28.74 (5.27)& 12.42 (3.11)\\ 
			ZDTS        & 6.24 (1.54) & 11.36 (1.84)& 17.68 (2.08) & 24.86 (1.65) & 22.48 (1.90)& 17.38 (3.10) \\
			\hline 
			
		\multicolumn{7}{p{14cm}}{\footnotesize \it Results are based on transformation of the $c_k$: 
			$\pi_{k}= (1+e^{-c_k})^{-1} - (1+e^{-c_{k-1}})^{-1}$ for $k=2, \ldots, K-1$ and $\pi_{1}=(1+e^{-c_1})^{-1}$ for $k=1$ for the ordered-multinomial and  of the $\mu$, $home$: $\lambda_1=e^{\mu+home}$,   $\lambda_2=e^\mu$ for the ZDTS models, respectively.}
		\end{tabular}
		\caption{ Posterior means of percentages ($\%$) of set-differences in a match between two competing teams of equal strength based on the ordered-multinomial and the ZDTS model; standard deviations ($\%$) are given within parentheses. }
		\label{probab-ordered}
	\end{table}
	
	\begin{comment}
			    	\hline
		Observed percentages & 0.18 & 0.17 & 0.10  & 0.10  & 0.20 & 0.27
			\end{comment}

\begin{figure}[h!]
		\centering
		\hspace{-1cm}
		\includegraphics[width=0.9\textwidth,keepaspectratio]{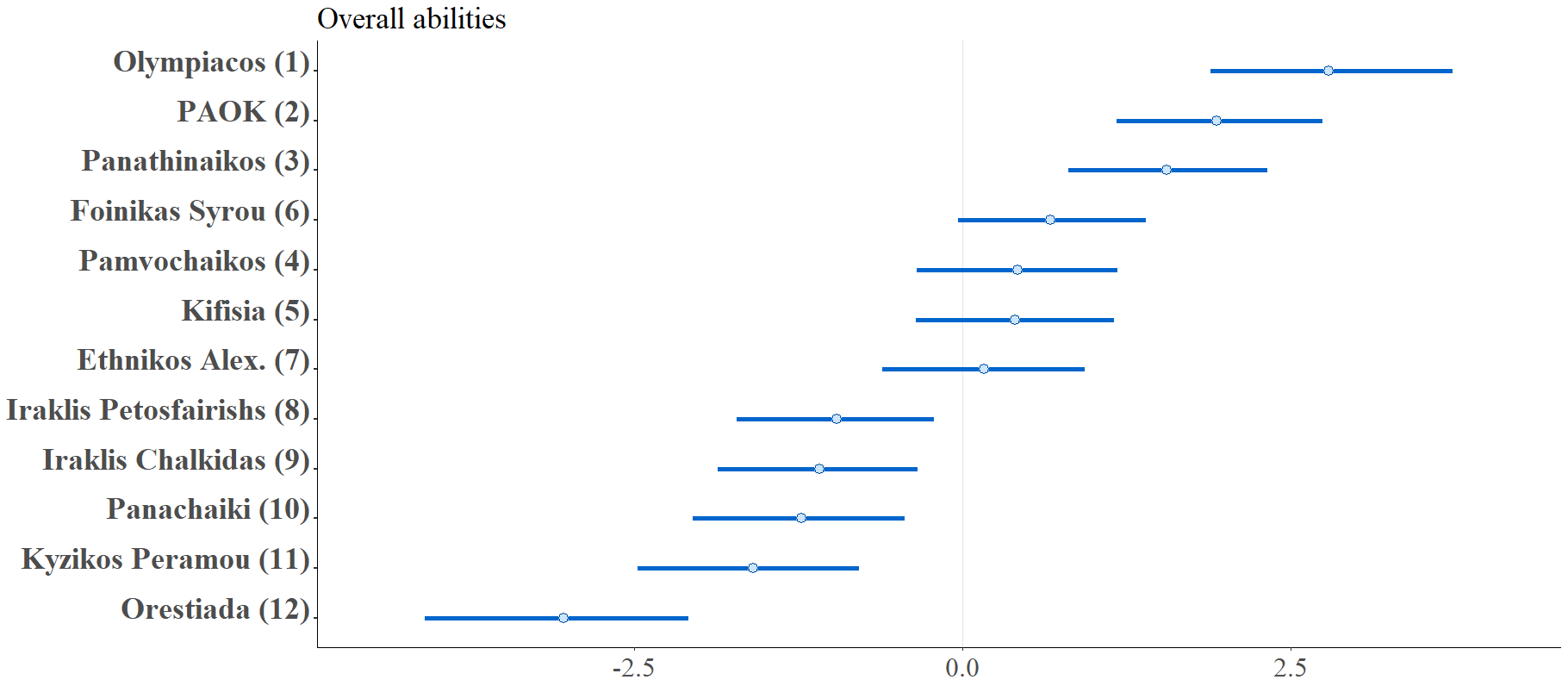}
		\centering
		\caption{95\% Posterior intervals of the ``net'' general ability of all teams for the ordered multinomial model (parameters $A_j$, $j=1,\dots,12$). The interval mid-points represent the posterior means; observed rankings are reported within parentheses.}
		\label{mul_ord_gen}
	\end{figure}

	\color{black}
	For the overall ability parameters of the ordered-multinomial model ($A_j$, for $j=1,\dots,12$), the $95\%$ posterior intervals are presented in 
	Figure \ref{mul_ord_gen} in descending order of the posterior mean. 
	Generally, the estimated abilities are in agreement with the final ranking of the league. 
	The only discrepancies we observe between the overall abilities and the final ranking positions concern the following teams: Foinikas Syrou (+2), Pamvochaikos (-1) and Kifisia (-1).

	Table \ref{probab-ordered} presents also the probabilities of set-differences for two teams of equal strength based the ZDTS model.  The picture is similar to the results of the ordered multinomial model, 
	but more weight is now given on the 3--2 difference which now has almost 25\% probability versus the 20\% observed with the ordered-multinomial model. Furthermore, the corresponding probabilities for differences of two and three sets in favour of the home team are almost equal to 22\%  and 17\%, respectively. Hence the ZDTS seems to suggest slightly higher values for the overall home effect. 
	
	Moreover, Table \ref{zdts_summaries} presents the posterior means of the constant term and the home effect both in the logarithmic and the original scale of sets. 
	From this table, we observe that the home effect increases  $\lambda_1$ against $\lambda_2$ by $41\%$. 
	%Although parameters $\lambda_1$ and $\lambda_2$ of the ZDTS are linked with the expected sets earned by the home and the away team, respectively, they do not directly reflect to proportional changes of the expected sets but to proportional changes of the expected latent unknown quantities (after removing a common  correlation component).

	Finally, we present the differences of the lambda parameters and the corresponding actual expected value when two teams of equal strength play to each other. 
	Concerning  the latter quantity, we observe that the posterior mean of the expected difference in sets is about 0.63 sets ranging from 0.19 to one set with posterior probability 95\%. 
	
	\begin{table}[h]
		\centering
		\begin{tabular}{l@{~~~~}l@{~~~~}c@{~~~~}c@{~~~~}cc@{~~}c}
			\hline
			&  &      &       &    & \multicolumn{2}{c} {\small Posterior Interval} \\  
			\cline{6-7} 
			Parameter Description &  & mean & median& sd &  $2.5\%$  & $97.5\%$ \\ 
			\hline
			Constant    (in log-sets)  & $\mu$      & 0.94 & 0.93 & 0.24 &  0.50 & 1.42 \\ 
			Home effect (in log-sets)  & $home$     & 0.34 & 0.34 & 0.12 &  0.11 & 0.61 \\ 
			Constant (in sets)         & $e^\mu$    & 2.64 & 2.54 & 0.64 &  1.64 & 4.16 \\ 
			Home effect ($\%$ in sets) & $e^{home}$ & 1.42 & 1.41 & 0.11 &  1.17 & 1.81 \\
			\hline 
			\\[-0.5em]
			\multicolumn{2}{l}{\underline{Estimated Quantities  for teams of equal strength$^\ddagger$}} \\  
			Parameter difference & $\lambda_1-\lambda_2$  & 1.08 & 1.03 & 0.47 & 0.29 & 2.14 \\ 
			Expected difference   & $E( Z_{ZDTS})$     & 0.63 & 0.63 & 0.22 & 0.19 & 1.04  \\
			\hline
			
		\end{tabular}
		\caption{Posterior summaries of constant $\mu$ as well as common home effect $home$ parameters and difference $\lambda_1-\lambda_2$ and the corressponding expected set-difference between two equal strength teams based on ZDTS model; $^\ddagger\lambda_1=e^{\mu+home}$ and  $\lambda_2=e^\mu$.}
		\label{zdts_summaries}
	\end{table}

\begin{figure}[h!]
		\vspace{-0.5cm} 
		\hspace{-0.5cm} 
		\includegraphics[width=0.95\textwidth,keepaspectratio]{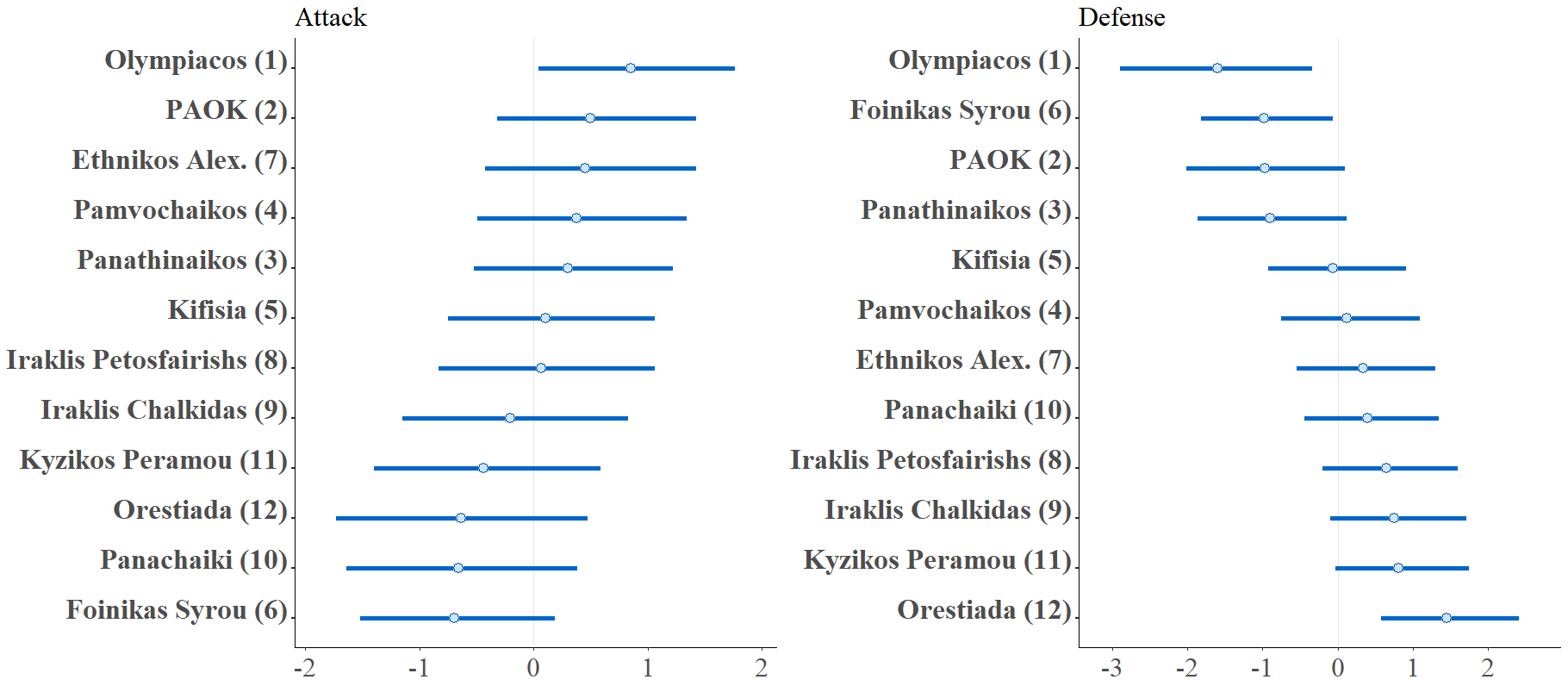}
		\caption{95\% Posterior intervals of both ``net" attacking and defensive parameters for the ZDTS model  (parameters $att_j$ and $def_j$ for $j=1,\dots,12$). The interval mid-points represent the posterior means;  observed rankings are reported within parentheses.}
		\label{zdts_att_def}
		
		\vspace{2em}
		\hspace{-2cm}  
		\includegraphics[width=0.8\textwidth,keepaspectratio]{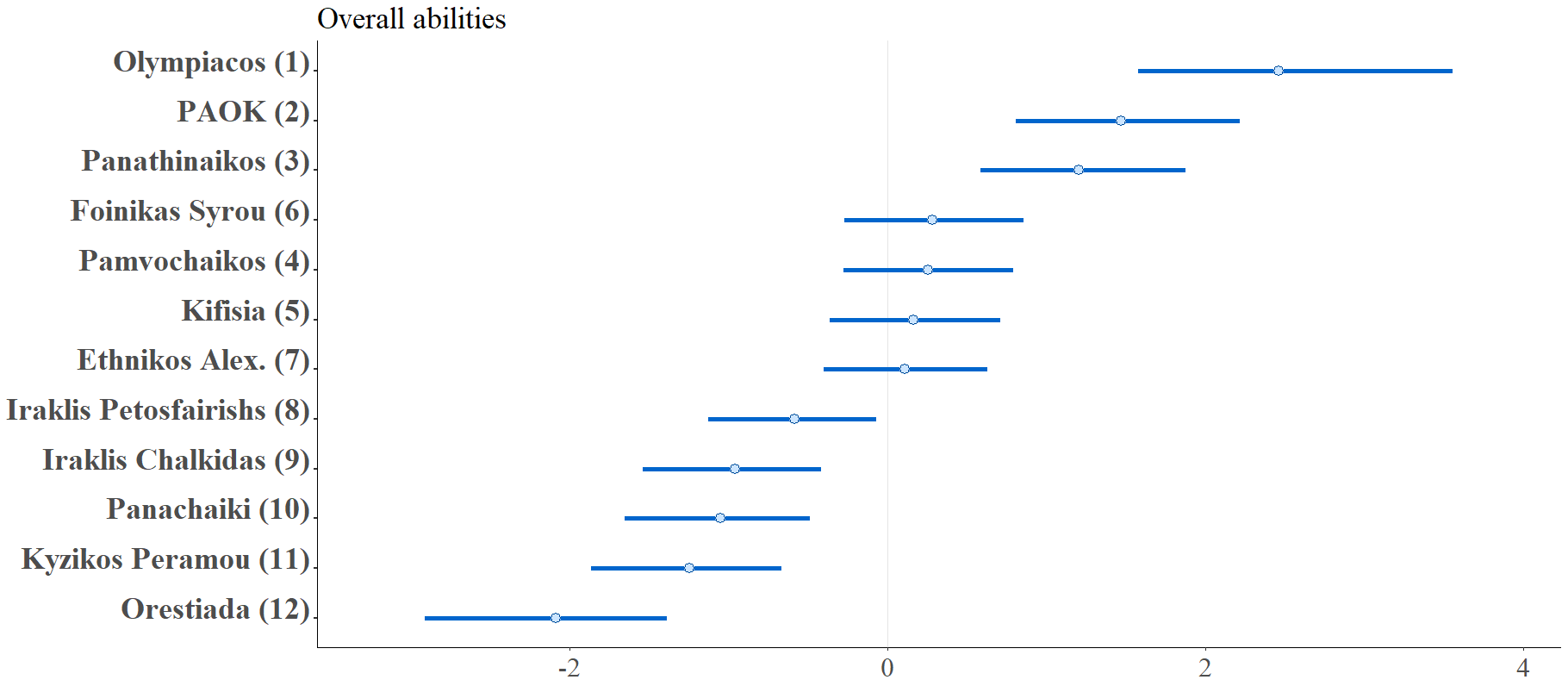}
		\centering
		\caption{95\% Posterior intervals of the overall team abilities ($o_j=att_j-def_j$) for all teams $j=1,\dots,12$ based on the ZDTS model. The interval mid-points represent the posterior means; observed rankings are reported within parentheses.}
		\label{zdts_overall_abil}
	\end{figure}

	Figure \ref{zdts_att_def} presents the 95\% posterior intervals for the attacking and defensive team abilities based on the ZDTS model, that is parameters $att_j$ and $def_j$ for $j=1,\dots, 12$. 
	We further present the differences between the attacking and defensive abilities for each team (calculated as $o_j=att_j-def_j$) which can play the role of overall team abilities and therefore they are comparable, in terms of interpretation, but with different scaling, to the corresponding abilities of the ordered multinomial model; see Figure \ref{zdts_overall_abil}. From this figure, it is obvious that the estimated overall abilities based on the ZDTS model have the  same behaviour with the corresponding abilities estimated by the ordered-multinomial model.
	Concerning the attacking and defensive abilities $att_j$ and $def_j$ (for $j=1,\dots,12$), 
	the picture is not the same since the final rankings and the ordering of these parameters do not coincide. The case of Olympiacos is notable since it is the best team concerning both the attacking and defensive abilities.  
	For many teams, their final ranking position is in agreement with either their attacking or their defensive abilities. For instance, the attacking ability of Pamvochaikos is in agreement with the corresponding final ranking position while its defensive ability is underestimated compared to the corresponding final ranking position. 
	A characteristic example of such discordance is Foinikas Syrou since it has the second best defensive ability but also the worst attacking ability. In essence, its excellent defensive performance compensates, in some extent, for its worst attack and for this reason the overall ability of this team is slightly overestimated (see also the final expected rankings in Section \ref{league_regen}).

	The parameters of the ZDTS are measuring the proportional differences on $\lambda_1$ and $\lambda_2$ 
which are linked with the mean of the ZDTS distribution via \eqref{mean_zdts} but they are not directly interpretable in terms of the expected set-difference  of the ZDTS model. 
Hence we have implemented an approximate analysis based on the posterior values of $\lambda_1$, $\lambda_2$ and 
$E(Z_{ZDTS})$ obtained using \eqref{mean_zdts} in order to understand the impact of the model parameters on the expected set-difference  of the ZDTS model. 
We have experimented with a variety of models in order to identify a simple but efficient association between  $E(Z_{ZDTS})$ and functions of $\lambda_1$ and $\lambda_2$ that would provide good approximating properties and easy to understand (and explain) relationship between these two quantities. 
We present results of a simple regression between the logarithm of $E(Z_{ZDTS})$ shifted by three units in order to ensure its positivity and the difference between $\log(\lambda_1)$ and  $\log(\lambda_2)$. 
The model was applied using two separate equations. One regression model was fitted for $\log(\lambda_1/\lambda_2) \le 0.8$ and another regression model was implemented for the rest of the values. The second case corresponds to matches where the home team is much better than the away team and hence we expect $E(Z_{ZDTS})$ to be close to a difference of three sets. Hence, small changes in the team ability of the home or the away team will have minor impact on the expected set-difference $E(Z_{ZDTS})$. 
The overall fit of the model is high with $R^2 = 0.96$. 
This approximate analysis provides an equation for the proportional change of the shifted expected set-difference which is given by
$$
\%\Delta \big[ E(Z_{ZDTS}) \big] = e^{\beta(b_1-b_2)}-1
$$
where $\beta$ is the correction effect obtained by the regression analysis of $E(Z_{ZDTS})$ on $\log(\lambda_1/\lambda_2)$ and $b_l$ is the effect of a covariate/ability on $\log \lambda_1$ and/or  $\log \lambda_2$, respectively.  
From this analysis, we have obtained that  $\beta=0.61$ for $\log(\lambda_1/\lambda_2) \le 0.8 $ and $\beta=0.07$ otherwise. 
Details about how these values are obtained are provided at Appendix C.2. 
In terms of our parameters, for the home effect we have $e^{0.61 \times (0.34-0.00)}=1.23$. 
This can be interpreted as a  $23\%$ increase in the expected shifted set-difference  (in favour   of the home team) for a match between two teams of equal strength when compared to the expected shifted set-difference in a neutral stadium (and hence no home effect is present). 
Similar is the interpretation for the ability parameters. For example, for Olympiacos the posterior means of the attacking and defensive parameters are equal to $0.85$ and $-1.61$. 
Hence, we expect that Olympiacos will have  shifted set-difference equal to $e^{0.61 \times( 0.85+1.61 )}= 4.5$ times as high as an average team playing against the same opponent when $\log (\lambda_1/\lambda_1) \le 0.8$. 
In a match that Olympiacos is dominating its opponent ($\log (\lambda_1/\lambda_2)>0.8$) then the expected shifted set-difference  will be increased by $19\%$ in comparison with the expected shifted difference obtained by a team of average ability when playing against the same opponent\footnote{This value is calculated by $[e^{0.07 \times( 0.85+1.61 )}-1] \times 100= 19\%$}. 
Similarly, for Kifisia\footnote{Both percentage increases have been calculated by $e^{0.61 \times( 0.10+0.06 )}=1.102$ and $e^{0.07 \times( 0.10+0.06 )}=1.011$}  
the expected shifted set-difference is increased by $10\%$ and $1\%$ for the two cases of $\log(\lambda_1/\lambda_2)$.

	\subsection{Predictive checking and league regeneration}

	\label{league_regen}
	One of the most interesting and challenging tasks in sports is the prediction of future matches. From the Bayesian perspective, the generation of the outcome of a future volleyball match $i'$ between a home and away team (denoted by  $ht_{i'}$ and $at_{i'}$) can be implemented via the posterior predictive distribution
	\begin{eqnarray} \label{pos_pred_dis}
	\begin{aligned}
	f\big({y}^{pred}_{i'}|\dn{y}\big)= & \int f\big({y}^{pred}_{i'}| ht_{i'}, at_{i'}, \dn{\theta}\big) f(\dn{\theta}|\dn{y}) d\dn{\theta} 
	= E_{\dn{\theta}| \dn{y}} \left[ f\big({y}^{pred}_{i'}|ht_{i'}, at_{i'}, \dn{\theta}\big) \right]&  \\
	\end{aligned}
	\end{eqnarray}
	where  ${y}^{pred}_{i'}$ is the future (predicted) set-difference for  match $i'$ where teams 
	$ht_{i'}$ and $at_{i'}$ compete with each other at the home stadium of the first team. 
	Moreover, $f\big({y}^{pred}_{i'}| ht_{i'}, at_{i'}, \dn{\theta}\big)$ denotes the model's sampling distribution of the future match $i'$ and $f(\dn{\theta}|\dn{y})$ the posterior distribution of observed data vector $\dn{y}$.
	From the above, it is obvious that the posterior predictive distribution is simply the posterior expectation of the probability function of the set-difference for future match $i'$. 
	
	We can generalize the above expression for $n'$ future matches by 
	\begin{eqnarray} \label{pos_pred_dis}
	\begin{aligned}
	f\big(\dn{y}^{pred}|\dn{y}\big)= & \int f\big(\dn{y}^{pred}| \dn{X}^{pred}, \dn{\theta}\big) f(\dn{\theta}|\dn{y}) d\dn{\theta} 
	= E_{\dn{\theta}| \dn{y}} \left[ f\big(\dn{y}^{pred}| \dn{X}^{pred}, \dn{\theta}\big) \right]&  \\
	%\propto & \int f(\dn{y}^{pred}|\dn{\theta}) f(\dn{y}|\dn{\theta}) f(\dn{\theta}) d\dn{\theta}&    
	\end{aligned}
	\end{eqnarray}
	where  $\dn{y}^{pred} = ( {y}_1^{pred}, \dots, {y}_{n'}^{pred} )^T$ is a vector of length $n'$ with elements the future (predicted) values, $n'$ is the number of future matches and $\dn{X}^{pred}$ contains the match specific information related with the future (to be predicted) matches. 
	In our case  $\dn{X}^{pred}$ simply contains the dummy variables indicating the home and away teams for each match.
	Moreover, $f(\dn{y}^{pred}|\dn{\theta})$ denotes the likelihood of the future values $\dn{y}^{pred}$ which is averaged over the posterior distribition $f(\dn{\theta}|\dn{y})$.
	Note that for the ordered-multinomial model (\ref{ordered-volleyball}), $\dn{\theta}=(c_{1}, \, \ldots, \, c_{K-1}, \, A_{1}, \, \ldots, \, A_{p})$ while for the ZDTS, the parameter vector is given by 
	$\dn{\theta}=(\mu, \, home, \, att_{1}, \, \ldots, \, att_{p}, \, def_{1},\, \ldots, \, def_{p})$. 
	
	Although the computation of the predictive distribution \eqref{pos_pred_dis} may look cumbersome due to the  (multivariate) integral involved in this equation, in practice it is straightforward to estimate it through MCMC methods. We simply introduce an additional step in our algorithm where we generate samples of the possible future outcomes $\dn{y}^{pred}$ from the model's sampling distribution $f(\dn{y}^{pred}|\dn{\theta}^{(t)})$ for the given parameter values $\dn{\theta}^{(t)}$ generated at the $t$ iteration of the MCMC algorithm. 
	Thus, in a similar manner as in \citeasnoun{kar.ntzouf.2008}, we obtain a sample from the predictive distribution for the ordered multinomial model by using the following steps: 
	\begin{itemize}
		\item Update the model parameters for the $t$ iteration of the MCMC algorithm \linebreak  
		$\dn{\pi}^{(t)}=\dn{\theta}^{(t)}=\big(c_{1}^{(t)}, \ldots, c_{K-1}^{(t)}, A_{1}^{(t)}, \ldots, A_{p}^{(t)}\big)$  
		by obtaining an additional MCMC iteration from \Stan. 
		 %using as initial values the parameter values $\dn{\theta}^{(t-1)}$ of the previous iteration.
		\item For $i'=1,\dots,n'$, calculate the probabilities of the multinomial distribution for future match $i'$
		$\dn{\pi}_{i'}^{(t)} = \big( \pi_{i1}^{(t)},\ldots, \pi_{iK}^{(t)} \big)$  as a function of $\dn{\theta}^{(t)}$ using 
		the inverse cumulative logit link functions of (\ref{ordered-volleyball}). \
		\item For $i'=1,\dots,n'$,  generate predictive values $y_{i'}^{rep}$ from the multinomial distribution 
		with probability parameters $\dn{\pi}_{i'}^{(t)}$, that is 
		$y_{i'}^{rep} \sim Multinomial \big( \dn{\pi}_{i'}^{(t)} \big)$. \ 
	\end{itemize}

	For the second modelling approach, since the ZDTS is not a standard distribution available in \Stan, 
	we have specified a user-defined distribution. Hence the generation of the predictive values can be summarized by the following steps:

	\begin{itemize}
		\item Update the model parameters for the $t$ iteration of the MCMC algorithm  
		$$\dn{\theta}^{(t)}= \big(\mu^{(t)}, home^{(t)}, att_{1}^{(t)},\ldots, att_{p}^{(t)}, def_{1}^{(t)},\ldots, def_{p}^{(t)}\big)$$ 
		by obtaining a single MCMC iteration from \Stan using as initial values the parameter values $\dn{\theta}^{(t-1)}$ of the previous iteration.
		
		\item For $i'=1,\dots,n'$,  calculate the parameters of ZDTS by 
		$\lambda_{i'1}^{(t)}= \exp{ \Big(\mu^{(t)}+home^{(t)}+att_{ht_{i'}}^{(t)}+def_{at_{i'}}^{(t)}\Big)}$ and $\lambda_{i'2}^{(t)}= \exp{ \Big(\mu^{(t)}           +att_{at_{i'}}^{(t)}+def_{ht_{i'}}^{(t)}\Big)}$. 
		
		\item For $i'=1,\dots,n'$ and $k=1,\dots, 6$, calculate $\pi_{i'k}$ from 
		\begin{equation}
		\pi_{i'k}^{(t)}=\frac{f_{Sk}\big( z_k |\lambda_{i'1}^{(t)}, \lambda_{i'2}^{(t)}\big)}
		{  \sum \limits_{z \in \{1,2,3\}}\ \Big\{ 
			f_{Sk}\big(-z| \lambda_{i'1}^{(t)}, \lambda_{i'2}^{(t)}\big) + 
			f_{Sk}\big( z| \lambda_{i'1}^{(t)}, \lambda_{i'2}^{(t)}\big) \Big\}}, 
		\label{ZDTS_multiprobs}
		\end{equation}
		where $z_k = k-3-\mathcal{I}( k\le 3 ) $.
		
		\item For $i'=1,\dots, n'$, generate $w_{i'}$ from a multinomial distribution with parameter vector $\dn{\pi}_{i'}^{(t)}$ where each element is calculated by \eqref{ZDTS_multiprobs}. 
		Hence,  $w_{i'}  \sim Multinomial \big( \dn{\pi}_{i'}^{(t)} \big)$.

		\item For $i'=1,\dots, n'$, set $y_{i'}^{rep} =w_{i'}-3-\mathcal{I}( w_{i'}\le 3 ) $. 
	\end{itemize}

	Through the predictive distribution, we can also regenerate the league standings which is a valuable tool for the evaluation of the goodness of fit of the implemented models. 
	This approach was introduced by \citeasnoun{lee1997} within the classical frequentist approach and it  was further adopted by \citeasnoun{kar.ntzouf.2008} for the simulation of the final league ranking table within the Bayesian framework.  In essence, a model has a satisfactory fit to the data when the ranking from its posterior predictive distribution is in general agreement with the observed ranking. 
	Thus, the model with the smallest discrepancies between the observed and predicted ranking is the one with the best fit in terms of the final league ranking. 
	
	In essence, we can obtain the predictive distribution of the ranking table by generating each match outcome of the whole regular season according to the previously mentioned procedure for the prediction of match outcomes. 
	Once we obtain one set of predicted matches for the whole season (which corresponds to one iteration in the MCMC algorithm), we can construct the final league table and its corresponding summary statistics (rankings, total points, total goals scored etc.). 
	Repeating this over all MCMC iterations, will give us a sample from the league statistic of interest (for example from the rankings) as an output. Using this method, we can construct the predictive final league ranking distribution. 
	
	The key feature here is that we generate predictions for the matches that we have already observed. 
	Hence $n'=n$ and $\dn{X}^{pred}$ is exactly the same with the pre-game information available for the observed matches. 
	Moreover, each $y_i^{rep}$ now corresponds to the ``predicted'' set-difference for the observed match $i$ with observed set-difference $y_i$. 
	Having generated  ${\dn{y}^{pred}}^{(1)}, \dots,{\dn{y}^{pred}}^{(T)}$ as a sample of size $T$  from the predictive distribution $f({\dn{y}^{pred}} | \dn{y} )$, then we obtain a sample from re-generated leagues by the following procedure: 
	\begin{itemize}
		\item [~] \hspace{-1.2em} For every iteration $t \in \{ 1, \dots, T\}$
		
		\item Calculate 
		\begin{eqnarray*} 
			\psi_{i1}^{(t)} &=& 3 \times \mathcal{I} \big({y_i^{pred}}^{(t)} \ge 2 \big) + 2\times  \mathcal{I}\big({y_i^{pred}}^{(t)} = 1 \big)+ \mathcal{I}\big({y_i^{pred}}^{(t)} = -1 \big), \mbox{~and} \\ 
			\psi_{i2}^{(t)} &=& 3 \times \mathcal{I}\big({y_i^{pred}}^{(t)} \le -2 \big) + 2 \times  \mathcal{I}\big({y_i^{pred}}^{(t)} = -1 \big)+ \mathcal{I}\big({y_i^{pred}}^{(t)} = 1 \big), 
		\end{eqnarray*} 
		where $\mathcal{I}(A)$ is the indicator function which takes the value of one if statement $A$ is true and zero otherwise. 			 
		
		\item For $j=1,\dots,p$, calculate the points of each team $j$ at league (or iteration) $t$ by 
		$$
		P_{j}^{(t)} =  \sum _{i=1}^n  \left\{ \psi_{i1}^{(t)} \times \mathcal{I}\big( ht_i = j \big) 
		+\psi_{i2}^{(t)} \times \mathcal{I}\big( at_i = j \big) \right\}.
		$$
		\item For $j=1,\dots,p$, obtain the final ranking $R_{j}^{(t)}$  (in descending order) of each team $j$ at league (or iteration) $t$ 
		that corresponds to the final set of points $P_1^{(t)}, \dots, P_p^{(t)}$. 
		
	\end{itemize}
	
	Then, the posterior predictive distribution of both league points and ranking of each team are readily available. Here, for the goodness-of-fit purposes, we use the posterior means of the points but also more thorough analysis can be based on the frequency distribution of the final ranking of each team; see for examples in \cite{kar.ntzouf.2008}.
	
	\subsection{Goodness-of-fit model comparison}
	\label{section_gof}

\begin{figure}[h!]
	\includegraphics[width=1\textwidth,height=0.8\textheight,keepaspectratio]{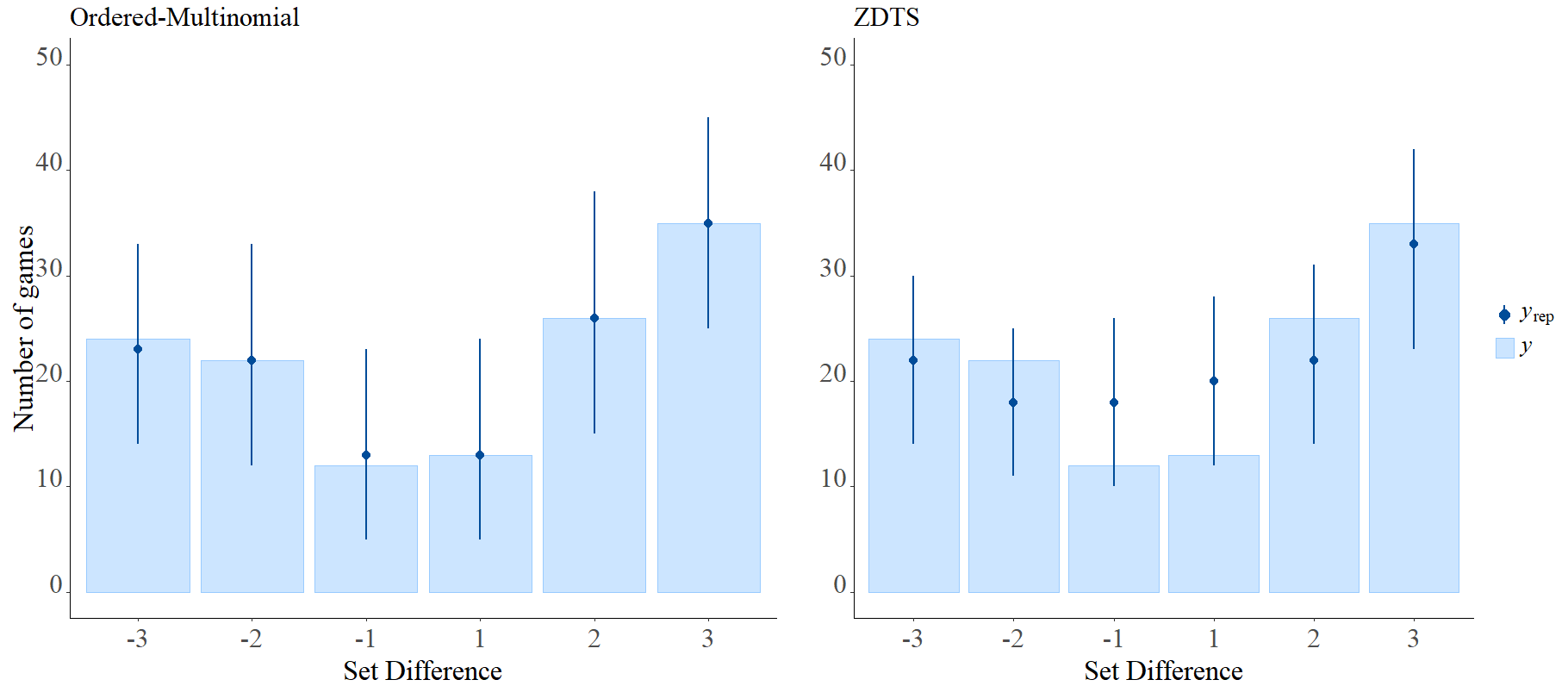}
	\caption{95\% Posterior intervals of predicted frequencies of set-differences; $y_{rep}$ and $y$ are the generated (median) and observed quantities, respectively.}
	\label{ord_zdts_barplots}
\end{figure}

	Figure \ref{ord_zdts_barplots} presents  the 95\% posterior intervals of the predicted frequencies of set-differences compared with the corresponding observed ones. Both models provide good fit to the data since the posterior medians (dark points) are   close to the observed frequencies. The 95\% posterior intervals in both models include the corresponding observed ones for each set-difference. The ordered-multinomial model seems to describe all the observed set-differences more accurately than the ZDTS model. 
	Concerning the ZDTS model, small differences are observed between the posterior medians and the observed frequencies for values of $|Z|\ge2$. For $|Z|=1$ (i.e., for tie-break matches), these differences are greater and the observed frequencies are marginally included in the reported posterior intervals. 
	The predictive performance of the ordered-multinomial model, with respect to the marginal frequencies of the differences, was   expected to be better than the   corresponding one for the ZDTS model since this first model focuses on the estimation of the probabilities (and consequently the frequencies) of set-differences rather than the estimation of the expected differences which is the focus of the latter model.

	Summaries of the posterior predictive league table for both models are presented in Table \ref{Ranking}. 
	The posterior mean points and the predicted rankings (based on the posterior mean points) for each team  along with the corresponding observed points and rankings are presented. 
	Predicted and observed rankings are in general agreement. 
	Only in positions 4--6 we can observe a mismatch   between the predicted and the observed rankings.
	%According to this table, the only differences between the predicted and the observed ranking concern positions 4--6. 
	More specifically, Foinikas Syrou was overestimated while Pamvochaikos was underestimated by both models. 
	Foinikas Syrou has gained three and one point more than the expected ones (ranked to position 4) according to the ordered-multinomial and the ZDTS models, respectively. 
	On the contrary, Pamvochaikos was predicted to have four and one point lower than the observed ones ranked in positions 6 and 5 for the two models, respectively. 
	Finally, the points of Kifisia were predicted accurately by the ordered-multinomial model (ranked correctly at position 5) while the ZTDS model predicted one point less than the observed ones (ranked at position 6 instead of the 5th). 
	Overall, when focusing on expected points, it seems that the ZDTS is more accurate than the ordered-multinomial. 
	The differences in terms of overall league points earned by each team, are even more obvious for teams in positions 8--9 where the absolute difference between their predicted and observed points is greater than five points for the ordered-multinomial while it is about three points for the ZDTS model. 
	
	\begin{table}[ht]
		\centering
		\small 
		\begin{tabular}{@{}r@{~}l@{~}l@{}r@{~\,}l@{~}|r@{~~}l@{~}l@{}r@{~\,}l@{}}
			\toprule
			\multicolumn{10}{c}{Model} \\
			\multicolumn{5}{c}{Ordered-Multinomial} & \multicolumn{5}{c}{Zero-Deinflated \& Truncated Skellam (ZDTS)}\\ 
			\midrule
			\midrule	
			\multicolumn{2}{c}{Ranking} & & \multicolumn{2}{c|}{Points} &
			\multicolumn{2}{c}{Ranking} & & \multicolumn{2}{c}{Points} \\ 
			Pred.$^\dagger$ & ~(Obs.)$^\ddagger$ & Teams & Post.Mean & ~(Obs.)$^\ddagger$ & \multicolumn{1}{|c@{}}{
				Pred.$^\dagger$}& ~(Obs.)$^\ddagger$ & Teams & Post.Mean & ~(Obs.)$^\ddagger$ \\
			\midrule
			1 & (1) & Olympiacos            & 58.87 &(62) &  1 &(1) & Olympiacos           & 59.98 &(62)\\
			2 & (2) & PAOK                  & 53.00 &(53) &  2 &(2) & PAOK                 & 52.97 &(53) \\
			3 & (3) & Panathinaikos         & 49.20 &(50) &  3 &(3) & Panathinaikos        & 49.87 &(50)\\
			4 & (6) & Foinikas Syrou        & 40.28 &(37) &  4 &(6) & Foinikas Syrou       & 38.04 &(37) \\
			5 & (5) & Kifisia               & 36.89 &(37) &  5 &(4) & Pamvochaikos         & 37.88 &(39) \\
			6 & (4) & Pamvochaikos          & 35.06 &(39) &  6 &(5) & Kifisia              & 36.17 &(37) \\
			7 & (7) & Ethnikos Alex.      & 34.81 &(36) &  7 &(7) & Ethnikos Alex.     & 35.64 &(36) \\
			8 & (8) & Iraklis Petosfairisis & 22.51 &(28) &  8 &(8) & Iraklis Petosfairisis& 24.88 &(28) \\
			9 & (9) & Iraklis Chalkidas     & 21.17 &(16) &  9 &(9) & Iraklis Chalkidas    & 19.54 &(16) \\
			10& (10) & Panachaiki            & 19.70 &(16) & 10 &(10)& Panachaiki           & 18.82 &(16) \\
			11& (11) & Kyzikos Peramou      & 16.34 &(14) & 11 &(11)& Kyzikos Peramou      & 15.67 &(14) \\
			12& (12) & Orestiada            & 05.70 & (7) & 12 &(12) & Orestiada           & 06.55 &(7) \\ \bottomrule
			\multicolumn{10}{p{13cm}}{\footnotesize \it $^\dagger$predictive ranking is according to the expected points; $^\ddagger$observed ranking and points are given within parentheses.}
		\end{tabular}
		\normalsize 
		\caption{Posterior predicted mean and observed points and rankings for the regular season data for both models with team abilities (Eq. \ref{ordered-volleyball} and \ref{ZDTSLIN}).}
		\label{Ranking}
	\end{table}
	
	In order to quantify the overall goodness-of-fit, we have calculated the mean absolute deviances (MAD) based on  the  predictive quantities of interest including results presented in Figure \ref{ord_zdts_barplots} and Table \ref{Ranking}.  
	The mean absolute deviance (MAD) for $t$ iteration is calculated by 
	\begin{equation}\label{Dev_meas_reg}
	%MAD\big(Q^{(t)}\big)=\frac{1}{\lvert Q\rvert}\sum_{j=1}^{\lvert Q\rvert} \Big| E\Big({Q_{j}^{pred}}^{(t)} \Big| \dn{y}\Big)-Q_{j} \Big|
	MAD\big(Q^{(t)}\big)=\frac{1}{\lvert Q\rvert}\sum_{j=1}^{\lvert Q\rvert} \Big| {Q_{j}^{pred}}^{(t)} -Q_{j} \Big|
	\end{equation}
	for $j= 1, \ldots, \lvert Q\rvert$ and $t=1, \ldots, T$ where $\lvert Q\rvert$ is the length of vector $Q$ on which the corresponding measure is based and $T$ the number of post--warmup MCMC iterations. 
	We have considered the following five different quantities as $Q$ in the calculation of MAD: 
	\begin{enumerate} 
		\item the number of predicted matches (frequencies) of each set-difference as given in Figure \ref{ord_zdts_barplots} ($|Q|=6$), 
		\item the proportion of matches (relative frequencies) of each set-difference corresponding to the results of Figure \ref{ord_zdts_barplots} ($|Q|=6$), 
		\item the set-differences ($Q_i=Z_i$ for $i=1,\dots,n$ and $|Q|=n=132$), 
		\item the total league  points of each team given in Table \ref{Ranking} ($Q_j=P_j$ for $j=1,\dots,p$ \& $|Q|=p=12$), and 
		\item the total set-differences of each team obtained from the final league table at the end of the regular season ($|Q|=p=12$). 
	\end{enumerate}

	The first two measures of Table \ref{dev_meas_in}  quantify and confirm the superiority of the ordered multinomial with respect to the marginal distribution of possible differences as already stressed from  Figure \ref{ord_zdts_barplots}. 
	However, ZDTS performs better in terms of overall points and set-differences per match indicating that it can regenerate the final league standings in a  more accurate way (measures 4 and 5). 
	Finally, in terms of overall set-differences for each team, the two models perform similarly with the ZDTS being slightly better with  an average absolute difference of about one set per team.  
	
	\begin{table}[h!]
		\centering
		\begin{tabular}{lccc}
			\hline \\[-0.5em] 
			\multicolumn{4}{c}{ \underline{Goodness of Fit Predictive Measures (in-sample diagnostics)} } \\[0.5em]  
			Comparison & $|Q|$ & \begin{tabular}[c]{@{}c@{}}Ordered\\ multinomial \end{tabular} &  ZDTS\\[0.5em]  \hline \\[-0.5em] 
			1. Frequencies of set-differences  & 6 & \textbf{4.11 (1.43)} & 4.99 (1.54) \\[0.5em] 
			2. Relative frequencies of set-differences ($\%$) & 6& \textbf{3.12 (1.08)} & 3.78 (1.17)\\[0.5em] 
			3. Expected set-difference       & 132& 1.46 (0.13)  & \textbf{1.39 (0.11)}  \\[0.5em] 
			4. Expected total points of each team  & 12 & 4.93 (1.17) & \textbf{4.14 (1.01)} \\[0.5em] 
			5. Expected total set-differences of each team & 12 & 7.83 (1.87) & \textbf{6.94 (1.68)} \\[0.5em]
			\hline
		\end{tabular}
		\caption{ Posterior means of Mean Absolute Deviance (MAD) between observed and predicted measures for the regular season; posterior standard deviations are given within parentheses.}
		\label{dev_meas_in}
	\end{table}
	
	\subsection{Out-of-sample prediction Analysis} 
	\label{out_of_sample}
	
	Section \ref{section_gof}  presents results from in-sample goodness-of-fit diagnostics based on the posterior predictive distribution. 
	In this section, we proceed further by evaluating the predictive performance of the proposed models in two out-of-sample scenarios. 
	In the first one, we use the data of the first half of the season in order to evaluate the perrfomance of the model for the second half of the season (mid-season scenario) while in the second scenario, 
	we examine the performance of our models in each play-off round using all previous data (play-offs prediction scenario).

	\subsubsection{Mid-season analysis.} 
	\label{midseason} 
	
	%Section \ref{section_gof}  presents results from in-sample goodness-of-fit diagnostics based on the posterior predictive distribution. 
	Here, we proceed further by evaluating the predictive performance of the proposed models by calculating the same measures as in Section \ref{section_gof}  in out-of-sample scenarios. 
	We focus on mid-season scenario where the results of the first half of the season are used for estimation (or learning) of the model parameters and the second half for testing the predictive performance of the models. 
	Hence we consider as the observed data $\dn{y}$ all set-differences $y_i$ for $i=1,\dots, 66$ (sample size equal to $n/2$) as the training observed dataset (along with the corresponding $\dn{X}_i$) while the test or predictive data $\dn{y}^{pred}$ has elements $y_i^{pred}=y_{i+66}$ for $i=1,\dots, 66$.

	According to Table \ref{dev_meas_out_half}, we surpisingly observe that the picture is reversed in comparison to the corresponding in-sample measures of Table \ref{dev_meas_in}. 
	Now the ordered-multinomial model is slightly better on the expected set-differences, on the total league points and on the total league set-differences while it is slightly worse on the marginal frequencies and proportions of set-differences. 
	
	\begin{table}[h!]
		\centering
		\begin{tabular}{lccc}
			\hline \\[-0.5em]
			\multicolumn{4}{c}{ \underline{Mid-Season Predictive Measures (out-of-sample diagnostics) }} \\[0.5em] 
			Comparison & $|Q|$ & \begin{tabular}[c]{@{}c@{}}Ordered\\ multinomial \end{tabular} &  ZDTS\\ \hline \\[-0.5em] 
			1. Frequencies of set-differences                 & 6 & 3.35 (1.04) & \textbf{3.13 (0.91)} \\[0.5em] 
			2. Relative frequencies of set-differences ($\%$) & 6 & 5.07 (1.58) & \textbf{4.75 (1.38)} \\[0.5em] 
			3. Expected set-difference                     &66& \textbf{1.57 (0.16)}  & 1.61 (0.15) \\[0.5em] 
			4. Expected total points of each team          & 12 & \textbf{3.35 (0.73)}  & 3.57 (0.75)\\[0.5em] 
			5. Expected total set-differences of each team & 12 & \textbf{6.14 (1.22)}  & 6.34 (1.27)\\[0.5em]
			\hline
		\end{tabular}
		\caption{Posterior means of Mean Absolute Deviance (MAD) between predicted and observed measures for the second half of regular season;within parentheses are their standard deviations.}
		\label{dev_meas_out_half}
	\end{table}

	%\subsection{Out-of-sample prediction: Play-offs analysis} 
	%\label{playoffs}
	
	%In this section  we evaluate the predictive performance of our models during the play-offs matches, which follow after the end of the regular season. 
	%We first focus on reporting the percentage of correct predictions with respect to the team which qualified to the next round followed by an analysis about the expected set-difference in each play-off round.

	\subsubsection{Play-offs analysis: Evaluation of the correct prediction of the qualified team.} \label{play-offs-qualification}
	
	We continue with the second prediciton scenarion, where  we evaluate the predictive performance of our models during the play-offs matches, which follow after the end of the regular season. 
	We first focus on reporting the percentage of correct predictions with respect to the team which qualified to the next round followed by an analysis about the expected set-difference in each play-off round.
	
In each play-off round, for each pair of competing teams, the team which reaches first a pre-specified/required number of wins $N_{req}$ in a sequence of matches is qualified to the next play-off round. Naturally, in such tournament rounds (and more generally in knock-out matches),  interest  mainly lies on the prediction of which team qualifies to the next round rather than the prediction of the set-difference. In the Greek league, the number of required wins is equal to two ($N_{req}=2$) for the Quarter--finals and equal to three ($N_{rew}=3$)  for  both  the Semi--finals and the Final. 
	Hence, for each play-off phase, the number of matches between two opponents is at least $N_{req}$  and at most $N_{seq}=2N_{req}-1$. 
	This means that the maximum number of matches that can be played between two opponents is equal to three ($N_{seq}=3$)  for the Quarter--finals and five ($N_{seq}=5$)  for  both  the Semi--finals and the Final. 

	For each MCMC iteration, we generate a sequence of matches from the predictive distribution until we  announce a qualifying team; see Algorithm \ref{algo1} for a detailed description. 
	An alternative, simpler way to identify the winner in each pair of opponent teams in Algorithm \ref{algo1} is to always generate $N_{seq}=2N_{req}-1$ matches  and then announce as the qualifying team the one with the highest number of wins. 
	Although, this approach is equivalent to the Algorithm \ref{algo1} in terms of results,  it is less efficient since we are required to generate a larger number of matches than the ones needed in order to announce the winner. 
	
	Each prediction has been performed at the beginning of the corresponding play-off round using the data of the regular season and the data from the previous play-off rounds. 
	Hence, we run our models using three different data splits (one for each play-off round). 
	For example, in the Semi--finals (second play-off round), the training set included all regular season matches and additionally the observed outcomes of the matches in the Quarter--finals (first play-off round).

\begin{algorithm}[p]
	\colorbox{gray!25}{\parbox{0.95\textwidth}{
			\SetKwInOut{Input}{Input}
			\SetKwInOut{Output}{Output}
			\Input{MCMC sample for the parameters of each model using the appropriate data split.}
			\Output{ $\mathbf{Q}$: MCMC sample of binary qualifying indicators $Q_j^{(t)}$  
				($j=1,\dots, N_{teams}$   and $t=1, \dots, T$) in the form of $T \times N_{teams}$ matrix; 
				where $N_{teams}$ is the number of teams participating in each play-off round.  	
			} 
			\For{$t = 1$ \KwTo $T$}{
				\For{$k = 1$ \KwTo $N_{pairs}$}{ 
					Set $W_1=0$ and $W_2=0$\; 
					\While{ $W_1<N_{req}$ {~\bf or~} $W_2<N_{req}$}{	
						Update which team plays at home ($ht$) and which as away team ($at$)   in each match in a sequence of matches\; 
						Update the parameters of the model accordingly \; 
						For  each match of the sequence between  opponent teams of $k$ pair: Generate the predicted set-difference $y^{pred}$   from the sampling distribution of the implemented model\; 
					\If{second team (team 2) plays at home}{Update $y^{pred} \Rightarrow y^{pred}=-y^{pred}$}  	\eIf{$y^{pred}>0$}{$W_1=W_1+1$}{$W_2=W_2+1$}  
					}
					\eIf{$W_1=N_{req}$}{$Q_{1}^{(t)}=1$; ~$Q_{2}^{(t)}=0$}{$Q_{1}^{(t)}=0$;~$Q_{2}^{(t)}=1$}  
				}				
			}
			{
				return $Q$\;
			}
			{\small \it \underline{Indexes}: \\ 
				\indent ~\hspace{2em} $t=,1, \dots T$; $T$: number of MCMC iterations; \\ 
				\indent ~\hspace{2em} $k=1,\dots, N_{pairs}$; $N_{pairs}$: number of pairs of opponents to be predicted; Each pair of opponents corresponds to a sequence of matches until $N_{req}$ wins are reached from one of the opponent teams.\\
				\indent ~\hspace{2em} $W_1,W_2$: number of wins for home and away teams of pair $k$, respectively; \\%in match $k$, respectively; \\
			}
		}			
		\caption{Volleyball stochastic play-offs prediction algorithm}
		\label{algo1} 	
	}	
\end{algorithm}
  
The final output of Algorithm \ref{algo1} is a matrix of MCMC generated binary indicators, 
one for each team indicating whether it qualifies or not in the next round. 
The posterior mean of these indicators provides an estimate of the posterior predictive probability for the qualification of each team  to the next play-off round. 
Then, the overall percentage of correct qualification predictions in each play-off round is provided in Table \ref{prop_qual_prob} by calculating the average of how many times we correctly predicted the team(s) qualified to the next round.

	Table \ref{prop_qual_prob} presents the average percentage of correct predictions   in terms of qualification to the next round. This was run separately for each round (using $T=6000$ MCMC iterations for three chains) by using the data of the full season and the data of the previous play-off round for both fitted models. 
	From this table,
	the ZDTS model demonstrates  slightly better predictive performance for both the Quarter and the Semi--finals than the ordered-multinomial one. However, in the Final stage, both models failed to predict the final winner since they indicated that Olympiacos should win three matches without loosing any (i.e. 3--0). 
	However,  PAOK eventually has won the title against all odds  (21.41\% and 10.99\% qualification probability of PAOK for ordered multinomial and ZDTS, respectively) winning three matches versus two matches won by Olympiakos. %(9 sets won for each team)%  
	This result was surprising  since Olympiacos was performing exceptionally both in the regular season and in the play-offs matches, winning easily all its competitors until the final phase.

	\begin{table}[h]
		\centering
		\begin{tabular}{lcc}
			\hline \\[-0.5em] 
			\multicolumn{3}{c}{ Play-offs Predictive Measures (out-of-sample diagnostics)} \\[0.5em]  
			Comparison  & \begin{tabular}[c]{@{}c@{}}Ordered\\ multinomial\end{tabular} &  \begin{tabular}[c]{@{}c@{}}ZDTS\\ \end{tabular} \\ \hline \\[-0.5em] 
			1. Quarter--finals  (1-8)   & 82.32\% & \textbf{86.53\%} \\ \vspace{0.5em}
			2. Semi--finals     (1-4)   & 84.75\% & \textbf{86.00\%} \\ \vspace{0.5em}
			3.  Finals         (1-2)   & ~\textbf{21.41}\%  & 10.99\% \\ 
			\hline
		\end{tabular}
		\caption{Percentages of correct predictions in each play-off round for both models with team abilities.}
		\label{prop_qual_prob}
	\end{table}

	\subsubsection{Play-offs analysis: Evaluation of the predictive performance in terms of set-differences.}
	\label{play-offs-set-difference}

				Here, we evaluate the predictive performance of two models under consideration by calculating the MAD of the set-differences  (in a similar manner as in Section \ref{midseason}) for each play-off round.
				The training sets used here are the same ones with the ones used in Section \ref{play-offs-qualification}. 
				Then, we have calculated the MAD measure by generating the predicted set-differences from the posterior predictive distribution for the sequence of matches occurred in each play-off round under evaluation. 
				
				The results of  Table \ref{dev_meas_out_playoffs} are quite similar to the ones of Table \ref{prop_qual_prob} since the ZDTS model provides better predictions of the expected set--differences in both Quarter and Semi--finals,  while in the Final stage both models fail to predict the final outcome in terms of the set--difference 
				(posterior mean of MAD $\approx$ 2.7 when the observed set-differences in the sequence of play-off matches between PAOK and Olympiakos were -3, 3, 1, -3, 2).

	\color{black}
  
  	\begin{table}[h!]
  	\centering
  	\begin{tabular}{lccc}
  		\hline \\[-0.5em]
  		\multicolumn{4}{c}{ \underline{Play--offs Predictive Measures (out-of-sample Expected set-difference) }} \\[0.5em] 
  		Comparison & $|Q|$ & \begin{tabular}[c]{@{}c@{}}Ordered\\ multinomial \end{tabular} &  ZDTS\\ \hline \\[-0.5em] 
  		1. Quarter--finals  (1-8)   &9& 1.40 (0.44) & \textbf{1.29 (0.39)} \\ \vspace{0.5em}
  		2. Semi--finals     (1-4)   & 6&1.60 (0.65) & \textbf{1.44 (0.57)} \\ \vspace{0.5em}
  		3.  Finals         (1-2)   & 5&2.72  (0.75)& \textbf{2.71 (0.63)} \\ 
  	\hline
  	\end{tabular}
  	\caption{Posterior means of Mean Absolute Deviance (MAD) between predicted and observed set-difference outcomes for each play-off round; within parentheses \color{black} are their standard deviations.}
  	\label{dev_meas_out_playoffs}
  \end{table}
  
  \color{black}
  
  \subsubsection{Final conclusions of out-of-sample analysis.}
  Overall, the results of the out-of-sample prediction exercise could not clearly separate two models in terms of predictive performance. 
  For the mid-season season analysis the ordered-multinomial model demonstrated better predictive performance for specific predictive diagnostics while for the play-offs procedure the ZDTS model  performed better than its competitor.  
  Hence, we may conclude that the two models provide similar predictions for this specific dataset. 
	\subsection{Bayesian model comparison via Bayesian predictive criteria}

	 We conclude our analysis by comparing both implemented models in terms of the Bayesian predictive information criteria of both WAIC and LOOIC.
	Another alternative would have been to use the Deviance Information Criterion \citep[DIC]{Spiegelhalter_etal_2002}. 
	WAIC is considered as an improved version of DIC and therefore we expect to be similar in terms of results; 
	hence the latter is omitted here. 
	The advantage of these criteria is that they approximately measure the predictive ability of each model based on the leave-one-out cross-validation procedure using the output of a single MCMC run. 
	 
	\begin{table}[h]
		\centering
		\begin{tabular}{lcc}
			\hline \\[-0.5em] 
			\multicolumn{3}{c}{ Bayesian Comparison via Information Criteria} \\[0.5em]  
			Comparison  & \begin{tabular}[c]{@{}c@{}}Ordered\\ multinomial\end{tabular} &  \begin{tabular}[c]{@{}c@{}}ZDTS\\ \end{tabular} \\ \hline \\[-0.5em] 
			WAIC  & 379,7 & 379,7 \\ \vspace{0.5em}
			LOOIC  & 379,9 & 380,4 \\ 
			\hline
		\end{tabular}
		\caption{Bayesian Model Comparison via the Information Criteria of WAIC--LOOIC.}
		\label{WAIC,LOOIC}
	\end{table}

	The two models are identical in terms of WAIC (379,7 for both models) and  LOOIC (379,9 for the ordered multinomial
	vs 380,4 for ZDTS). 
	These results are in agreement with our previous goodness-of-fit and out of sample analysis presented in Sections \ref{section_gof}--\ref{out_of_sample}.  
	Thus, we can safely reach the conclusion that both models provide similar predictions in volleyball match outcomes in terms of set-differences for the Greek league dataset used in this paper. 

	\color{black}
	 
	\section{Discussion}
	\label{Discussion}
	In the present paper, the ordered-multinomial and the zero-deflated and truncated Skellam (ZDTS) models   are proposed and implemented for the analysis of the set-difference in volleyball matches. 
	These two models are fitted in a dataset from the Greek volleyball league by using the common vanilla model structure previously adopted in other sports, especially in association football; see, for example, in \cite{kar.ntzouf.2003}. 
	We have performed several goodness-of-fit and predictive comparisons from which no clear winner between the two models is announced.
	Both of the proposed models re-construct the final league sufficiently well 
	with the ZDTS model to behave slightly better in some occasions  and the ordered multinomial in others.
	The only case in which the ZDTS model is slightly superior than the ordered multinomial is in the case of the prediction of both Quarter and Semi--finals.
	
	Our future research plans include the extension of the proposed models in order to embody more detailed  information about each match. 
	We intend to focus on the number of specific skill events such as the false execution of serves, the perfect execution of  passes, settings, different types of attacks, among others. 
	This project is currently in progress by the authors and it will be combined with Bayesian variable selection methods in order to identify which skills are important determinants of the final outcome.  
	Such models can be used either for understanding the importance of each skill  or for predicting the result of the next match. The two problems need different handling.
	For example the first one is more straightforward to implement since the final statistics and skill events for each match will be directly introduced in the linear predictor of each model. 
	This modelling approach can offer useful consulting tools for the manager of each team which will enhance his understanding about  which team characteristics should be strengthened via strategic decisions (type of training, selection of specific player, transfer a new player). 
	Moreover, these models can be also used for the calculation of xSets (in accordance to xGoals used in soccer). 
	This approach follows the work of \cite{Miskin et al} where they used Bayesian logistic regression models to evaluate the importance of each skill at the final outcome of win or loss  for each volleyball match. 
	In a different perspective,  \citeasnoun{Drikos et al}, \cite{Miskin et al} and \cite{Florence et al} used a Dirichlet-Multinomial Bayesian model in order to estimate the marginal probability of reaching to a point for each skill in a rally  within a match. 
	Similar approach, within the Markovian framework, has been implemented by \citeasnoun{Heiner et al} in womens' soccer matches.
	Nevertheless, if we are interested for building a prediction model, then the covariates should be related with information available before the match. In this case, we may  use as covariates the average skill statistics over all or some matches before the match of interest.

	Another direction for future research is the implementation of the proposed models in a wide variety of national volleyball leagues or international tournaments as in \citeasnoun{Tsokos et al} for association football. 
	This will help us  generalize the empirical findings of this work and evaluate the overall importance of the proposed models for  volleyball in general. 
	Moreover, it would be beneficial to implement meta-analysis techniques in order to specify the necessary weights of importance of each league. %According to a possible application of network meta-analysis method in our research, each league would obtain its weight depending on its importance and impact on our analysis.

	Finally, implementation of Bayesian model averaging techniques \citep{Hoeting et al} can also be a useful addition in the relevant literature. The implementation of this methodology will allow us  to  account for model uncertainty over each model prediction and reach safer conclusions about the importance of each skill/feature.
	%%%%%%%%%%%%%%%%%%%%%%%%%%%%%%
	\markboth{}{\rm }

	%\clearpage
	%\vspace*{6pt}
	%%%%%%%%%%%%%%%%%bibliography style
	%\markboth{C. E. POWELL}{\rm PARAMETER-FREE H(DIV) PRECONDITIONING}
	
	% *** Double blinded material
	%\section*{Acknowledgements}
	%This research is financed by the Research Centre of Athens University of Economics and Business,in the framework of the project entitled ``Original Scientific Publications 2019''.
	\color{black}
	
	%%%%%%%%%%%%%%%%%%%%%%%%
	%%%%%%%%%%%%%%%%%%%%%%%%%%%%%%%%%%%
	
\end{document}

\label{FirstPage}
	\pagestyle{headings}
	\markboth{}{\rm For the Special Issue on Mathematics in Sports}

	%%%%%%%%%%%%%%%%%
	%	\title{Bayesian Prediction of Volleyball Sets Using the Truncated Skellam and the Ordered Multinomial Models}
	\title{Electronic Appendix of \\ {\it ``Bayesian Models for Prediction of the Set-Difference in Volleyball''}}
	
	\author{ {\sc Ioannis Ntzoufras$^{*}$},{\sc Vasilis  Palaskas}\footnote{This work has been completed at Athens University of Economics and Business; Current affiliation: Fantasy Sports Interactive, Athens, Greece} and {\sc Sotiris Drikos} \\[2pt]
		AUEB Sports Analytics Group, Computational and Bayesian Statistics Lab, \\
		Department of Statistics, University of Economics and Business, \\
		76 Patission Street, 10434 Athens, Greece\\[6pt]
		{\rm $^{*}$Corresponding author: ntzoufras@aueb.gr}\\
		{\rm [For the Special Issue on Mathematics in Sports]}\vspace*{6pt}}
	%{\rm [Received on ]}\vspace*{6pt}}
	
	\maketitle

\appendix 
\renewcommand{\thesection}{Appendix \Alph{section}}
\renewcommand{\thesubsection}{\Alph{section}.\arabic{subsection}}
\renewcommand{\thetable}{\Alph{section}.\arabic{table}}
\renewcommand{\thefigure}{\Alph{section}.\arabic{figure}}
\renewcommand{\thefigure}{\Alph{section}.\arabic{figure}}
\renewcommand{\theequation}{\Alph{section}.\arabic{equation}}

\color{black}

In this electronic appendix, we provide supplementary material and details for 
%(a) 
the {\sf R} and {\sf Stan} programming codes used in the main paper, 
%(b) 
the MCMC convergence diagnostics, 
%(c) 
the posterior distributions of parameters 
and the predictive performance of the implemented models. 

\section{Code Instructions}

	All data and the {\sf R} and {\sf Stan}  code of the paper are available the paper's github repository\footnote{
		\href{https://github.com/Vasilis-Palaskas/Bayesian-Models-for-Prediction-of-the-Set-Difference-in-Volleyball}{\sf  https://github.com/Vasilis-Palaskas/Bayesian-Models-for-Prediction-of-the-Set-Difference-in-Volleyball} }. 
		%\footnote{This is the blinded version. The URL will be reported in the final unblinded version.}
	%\href{https://github.com/Vasilis-Palaskas/Bayesian-Models-for-Prediction-of-the-Set-Difference-in-Volleyball}{\sf   https://github.com/Vasilis-Palaskas/Bayesian-Models-for-Prediction-of-the-Set-Difference-in-Volleyball} 
	 %\url{https://github.com/Vasilis-Palaskas/Bayesian-Models-for-Prediction-of-the-Set-Difference-in-Volleyball} 
	Detailed description and instructions   about how to   run the {\sf R} codes and   fully reproduce the paper's results are available at the same URL.

\section{MCMC Convergence Diagnostics (Additional Material for Section 3.3)} 

	Here we provide MCMC convergence diagnostics of the MCMC output used in the corresponding paper. 
	More specifically,   we present the results of the Raftery and Lewis (1992) convergence diagnostic (Appendix \ref{App_B1}) along with graphical representations (trace plots, ergodic plots and auto-correlation plots) of the MCMC output for the two implemented models (Appendices \ref{App_B2} and \ref{App_B3}, respectively). 
	All tests and figures confirm that the MCMC algorithms have been converged.

	\subsection{Raftery and Lewis Diagnostic}
	\label{App_B1}

	The results of the Raftery and Lewis (1992) convergence diagnostic for the MCMC output used in main body of the article are presented in this section. 
	The aim of this diagnostic is to reach a pre-specified degree of
	accuracy   for  specific quantiles rather than the convergence of the mean. 
	More specifically, this diagnostic reports the following quantities:
	\begin{itemize}
		\item {\it Effective Sample Size ($n_{eff}$) } is  an approximation of the effective number of independent draws from the posterior distribution. In case of high auto-correlation, the effective sample size is considerably smaller than the total number of MCMC iterations.
		\item {\it Potential Scale Reduction Factor (Rhat)} is the ratio of the average variability of samples within each chain over the variance of the pooled samples across chains. If the chains of the fitted model have converged to a common target (posterior) distribution, then the values of specific diagnostics will be close to one.

		\item $M$ is the number of suggested burn-in iterations.
		
		\item $N$ is the total number of MCMC iterations that the chain should be run. 
		
		\item $N_{min}$ is the minimum number of iterations required to estimate the quantile of interest
		with the pre-specified accuracy under the assumption of independence. 
		\item $I$ is the dependence factor given by $I = \tfrac{N}{N_{min}}$ which indicates the relative
		increase of the total sample due to autocorrelation. 
	\end{itemize}
In the output of this diagnostic, the most relevant quantity of interest is $I$. 
If $I$ is equal to one, then the generated values can be considered as independent (or close enough to independence).
On the other hand, values greater than 5 often indicate a problematic behaviour; see, for details, in Best et al. (1996). 
In general $I$ can  be considered as a rough estimate of the required thinning interval.
Note that if the current number of generated observations $T$ is lower than $N_{min}$ then an error is printed and the remaining values are not estimated. 

All the results presented here are obtained using the default values of the diagnostic in {\sf CODA}  {\sf R} package. 
From Tables \ref{RL_ordered} and \ref{RL_ZDTS}, it is obvious that both chains have been converged.

For the ordered multinomial  model, the final sample was obtained by running  3 MCMC chains. 
Each chain was run for 16000 iterations for which a thinning equal to 2 was applied, and then a burnin of 
$B=2000$ was eliminated (from the thinned chain) resulting to a final sample of 6000 iterations. 
In total $3\times 6000 = 18000$ iterations were kept. 
For the ZDTS multinomial the final sample was obtained by running   3 MCMC chains each one of final length of 6000 iterations. Each chain was obtained from runs of length equal to 40000 for a thinning equal to 5 was applied and then a burn-in of 
$B=2000$ was eliminated (from the thinned chain) resulting to a final sample of 6000 iterations   as in the ordered-multinomial MCMC runs.

	% Merged Chains
	
	\begin{table}[h]
		\centering
		\begin{tabular}{lcccccc}
			\hline
			& $n_{eff}$ & Rhat & M & N & Nmin & I \\ 
			\hline
			Constant 1   & 15318 & 1.00 & 2 & 3795 & 3746 & 1 \\ 
			Constant 2   & 15845 & 1.00 & 1 & 3743 & 3746 & 1 \\ 
			Constant 3   & 16115 & 1.00 & 2 & 3779 & 3746 & 1 \\ 
			Constant 4   & 16893 & 1.00 & 2 & 3693 & 3746 & 1 \\ 
			Constant 5   & 16722 & 1.00 & 2 & 3848 & 3746 & 1 \\ 
			Ethnikos Alexandroupolis & 14186 & 1.00 & 2 & 3795 & 3746 & 1 \\ 
			Foinikas Syrou & 13922 & 1.00 & 2 & 3710 & 3746 & 1 \\ 
			Iraklis Chalkidas & 14439 & 1.00 & 1 & 3743 & 3746 & 1 \\ 
			Iraklis Petosfairishs & 13527 & 1.00 & 2 & 3795 & 3746 & 1 \\ 
			Kifisia & 14558 & 1.00 & 2 & 3778 & 3746 & 1 \\ 
			Kyzikos Peramou & 14041 & 1.00 & 2 & 3727 & 3746 & 1 \\ 
			Olympiacos & 13676 & 1.00 & 2 & 3660 & 3746 & 1 \\ 
			Orestiada & 14158 & 1.00 & 2 & 3778 & 3746 & 1 \\ 
			Pamvochaikos & 13947 & 1.00 & 2 & 3795 & 3746 & 1 \\ 
			Panachaiki & 14246 & 1.00 & 2 & 3813 & 3746 & 1 \\ 
			Panathinaikos & 13920 & 1.00 & 2 & 3676 & 3746 & 1 \\ 
			PAOK & 17699 & 1.00 & 2 & 3710 & 3746 & 1 \\ 
			\hline
			\multicolumn{7}{p{10cm}}{ \small \it \textbf{MCMC output details:} chains=3; 
				For each chain: total length=$16000$ iterations, Burnin/warmup iterations $B=2000$, thin=2, iterations kept $T=6000$; Total number of iterations finally used=$18000$ iterations ($3\times 6000$)}
			
		\end{tabular}
		\caption{Raftery-Lewis   diagnostic values for the ordered-multinomial model.} 
		\label{RL_ordered}
	\end{table}

	\begin{table}[h]
		\centering
		\begin{tabular}{lcccccc}
			\hline
			& n\_eff & Rhat & M & N & Nmin & I \\ 
			\hline
			Constant $\mu$ & 18338 & 1.00 & 1 & 3743 & 3746 & 1 \\ 
			Home effect & 16411 & 1.00 & 1 & 3743 & 3746 & 1 \\ 
			\hline 
			Attacking Effects & \\ 
			\hline 
			Ethnikos Alexandroupolis  & 17292 & 1.00 & 2 & 3693 & 3746 & 1 \\ 
			Foinikas Syrou  & 17837 & 1.00 & 1 & 3743 & 3746 & 1 \\ 
			Iraklis Chalkidas  & 17451 & 1.00 & 1 & 3743 & 3746 & 1 \\ 
			Iraklis Petosfairishs  & 17744 & 1.00 & 2 & 3830 & 3746 & 1 \\ 
			Kifisia  & 17980 & 1.00 & 2 & 3676 & 3746 & 1 \\ 
			Kyzikos Peramou  & 17622 & 1.00 & 2 & 3813 & 3746 & 1 \\ 
			Olympiacos  & 17366 & 1.00 & 2 & 3710 & 3746 & 1 \\ 
			Orestiada  & 17610 & 1.00 & 2 & 3830 & 3746 & 1 \\ 
			Pamvochaikos  & 18086 & 1.00 & 2 & 3710 & 3746 & 1 \\ 
			Panachaiki  & 17829 & 1.00 & 2 & 3761 & 3746 & 1 \\ 
			Panathinaikos  & 17855 & 1.00 & 1 & 3743 & 3746 & 1 \\ 
			PAOK  & 18295 & 1.00 & 2 & 3830 & 3746 & 1 \\ 
			\hline 
			Defensive Effects & \\ 
			\hline 
			Ethnikos Alexandroupolis  & 17477 & 1.00 & 2 & 3813 & 3746 & 1 \\ 
			Foinikas Syrou  & 17850 & 1.00 & 2 & 3660 & 3746 & 1 \\ 
			Iraklis Chalkidas  & 17758 & 1.00 & 2 & 3693 & 3746 & 1 \\ 
			Iraklis Petosfairishs  & 17595 & 1.00 & 1 & 3743 & 3746 & 1 \\ 
			Kifisia  & 18177 & 1.00 & 2 & 3710 & 3746 & 1 \\ 
			Kyzikos Peramou  & 17714 & 1.00 & 2 & 3761 & 3746 & 1 \\ 
			Olympiacos  & 17703 & 1.00 & 2 & 3778 & 3746 & 1 \\ 
			Orestiada  & 17119 & 1.00 & 2 & 3778 & 3746 & 1 \\ 
			Pamvochaikos  & 18432 & 1.00 & 2 & 3778 & 3746 & 1 \\ 
			Panachaiki  & 18028 & 1.00 & 2 & 3710 & 3746 & 1 \\ 
			Panathinaikos  & 17279 & 1.00 & 1 & 3743 & 3746 & 1 \\ 
			PAOK  & 18319 & 1.00 & 2 & 3693 & 3746 & 1 \\ 
			\hline
			\multicolumn{7}{p{10cm}}{ \small \it \textbf{MCMC output details:} chains=3; 
				For each chain: total length=$40000$ iterations, Burnin/warmup iterations $B=2000$, thin=5, iterations kept $T=6000$; Total number of iterations finally used=$18000$ iterations ($3\times 6000$)}
		\end{tabular}
		\caption{Raftery-Lewis  diagnostic  values for the ZDTS model.} 
		\label{RL_ZDTS}
	\end{table}
		
	\clearpage
	\newpage 

	\subsection{Convergence Plots for the Ordered Logistic Model (chains=3, thin=2, Iterations $T=18000$, burnin $B=2000$ for each chain)}
	\label{App_B2}

	\begin{figure}[h!]
		\includegraphics[width=1\textwidth,height=0.8\textheight,keepaspectratio]{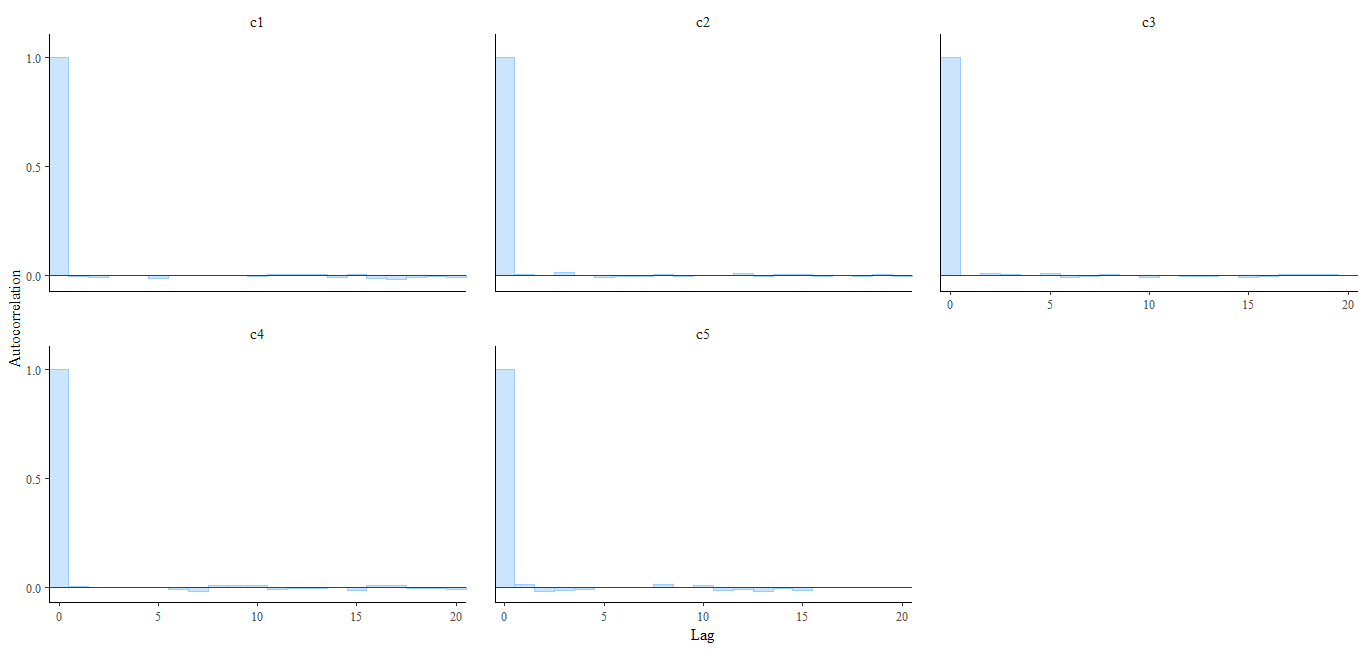}
		\caption{ACF plots for the constant parameters of the Ordered Multinomial Logistic.}
		\label{ord_acf_c_chains3}
	\end{figure} 
	
	\begin{figure}[h!]
		\includegraphics[width=1\textwidth,height=0.8\textheight,keepaspectratio]{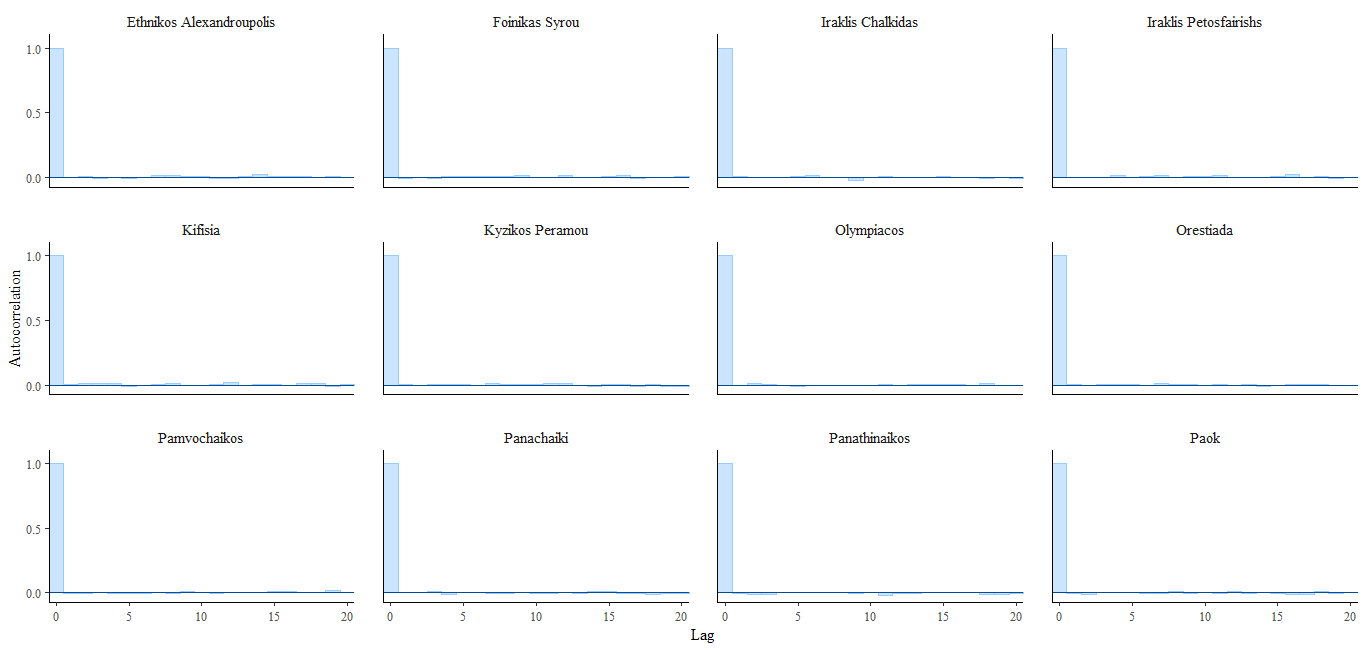}
		\caption{ACF plots for the team abilities parameters of the ordered-multinomial model.}
		\label{ord_acf_teams_chains3}
	\end{figure}

	%Ergodic
	\clearpage
	
	\begin{figure}[h!]
		\includegraphics[width=1\textwidth,height=0.8\textheight,keepaspectratio]{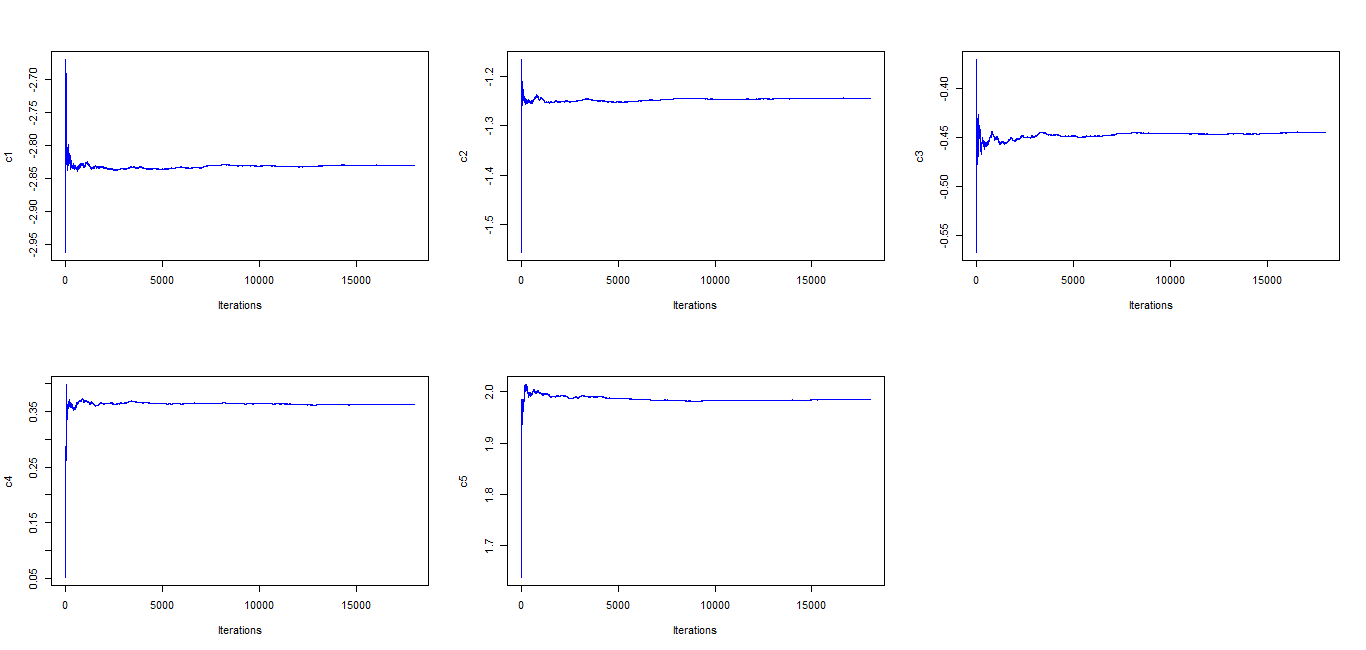}
		\caption{Ergodic plots for the constant parameters of the ordered-multinomial model.}
		\label{ord_erg_c_chains3}
	\end{figure} 
	
	\begin{figure}[h!]
		\includegraphics[width=1\textwidth,height=0.8\textheight,keepaspectratio]{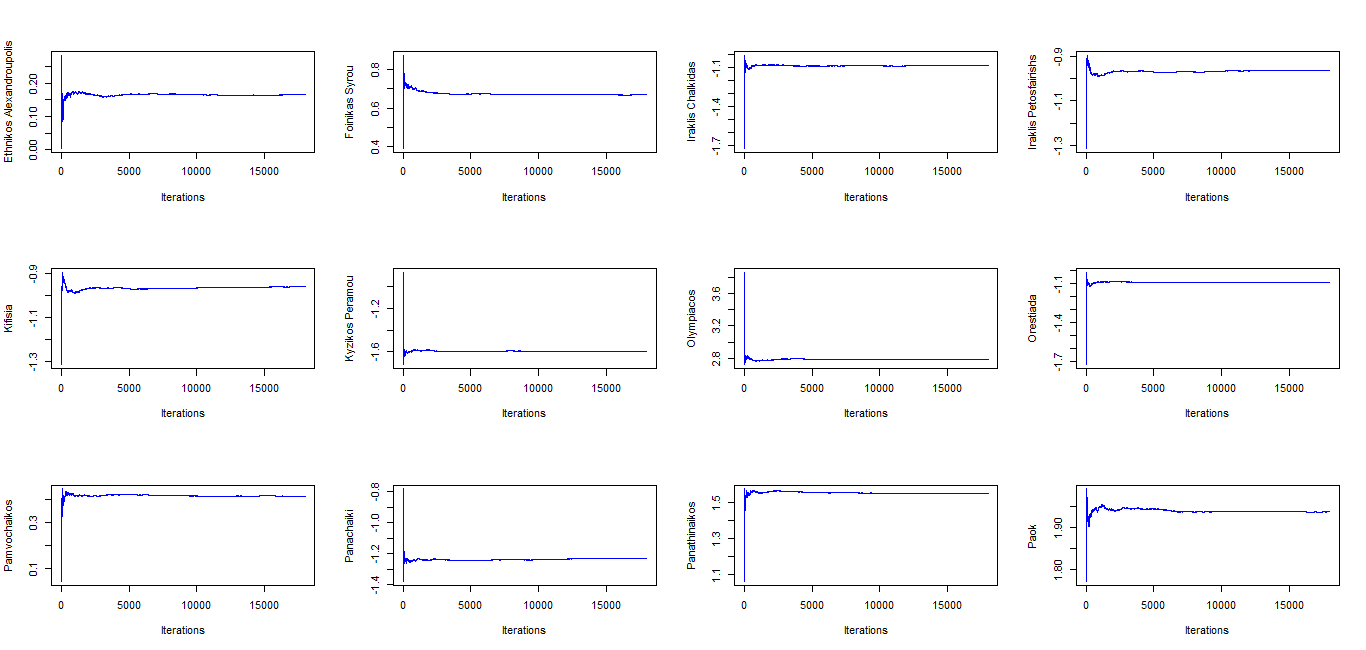}
		\caption{Ergodic plots for the team abilities parameters of the ordered-multinomial model.}
		\label{ord_erg_teams_chains3}
	\end{figure} 
	
	%Trace
	\clearpage
	
	\begin{figure}[h!]
		\includegraphics[width=1\textwidth,height=0.8\textheight,keepaspectratio]{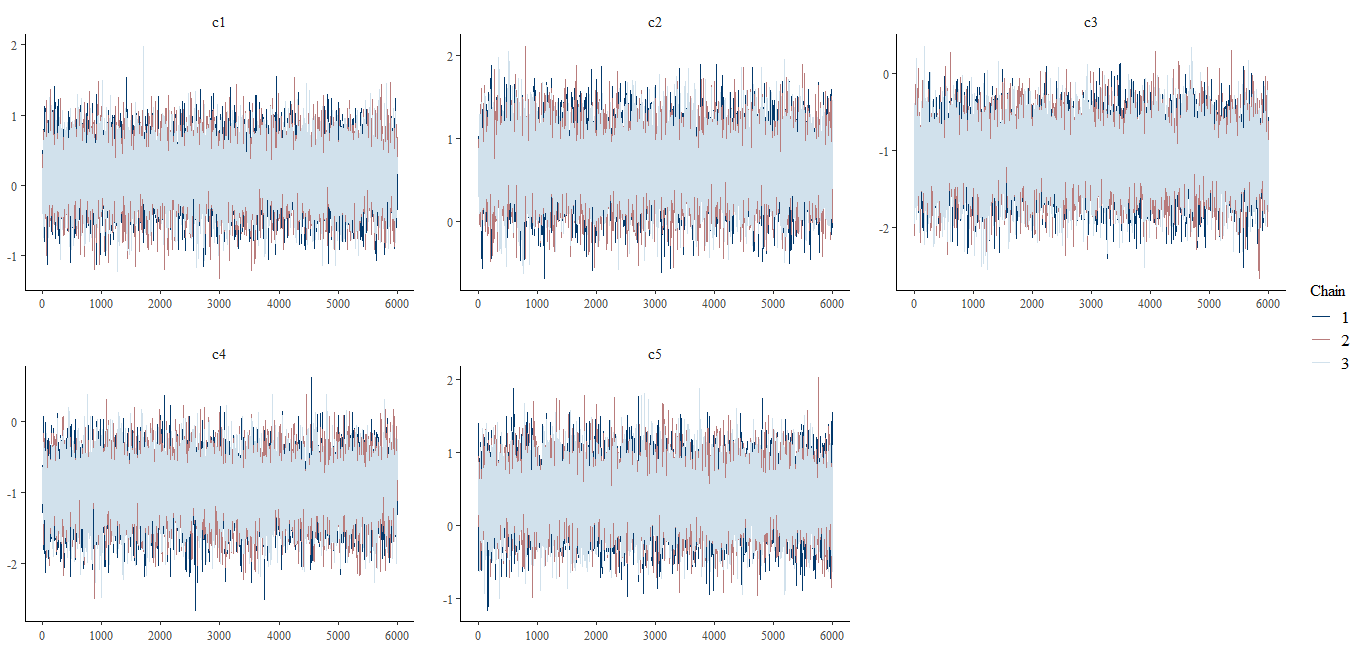}
		\caption{Trace plots for the constant parameters of the the ordered-multinomial model.}
		\label{ord_trace_c_chains3}
	\end{figure} 
	
	\begin{figure}[h!]
		\includegraphics[width=1\textwidth,height=0.8\textheight,keepaspectratio]{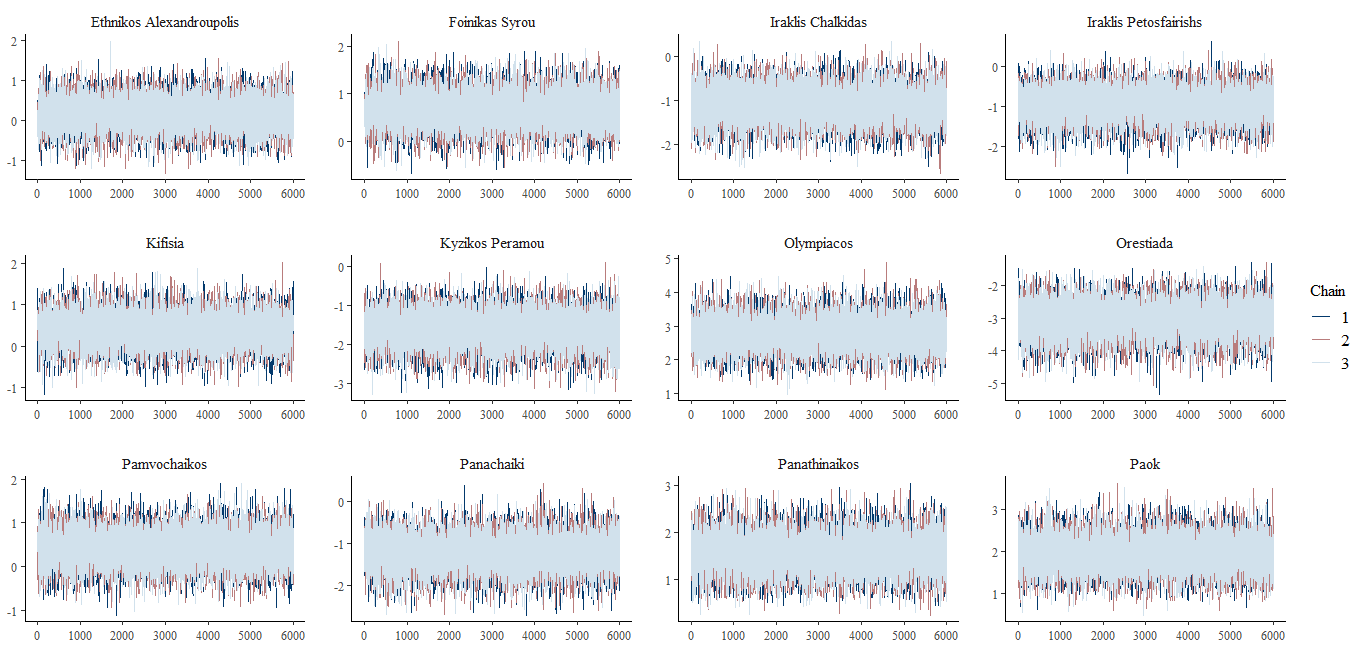}
		\caption{Trace plots for the team abilities parameters of the ordered-multinomial model.}
		\label{ord_trace_teams_chains3}
	\end{figure}

	\clearpage
	%Ordered Logistic model with chains=3, thin=2 (Final Version)

	%\clearpage
	
	\subsection{Convergence Plots for the ZDTS Model (chains=3, thin=5, Iterations $T=18000$, Burnin $B=2000$ for each chain)}
	\label{App_B3}

	\begin{figure}[h!]
		\begin{center} 
		\includegraphics[width=0.5\textwidth,height=0.8\textheight,keepaspectratio]{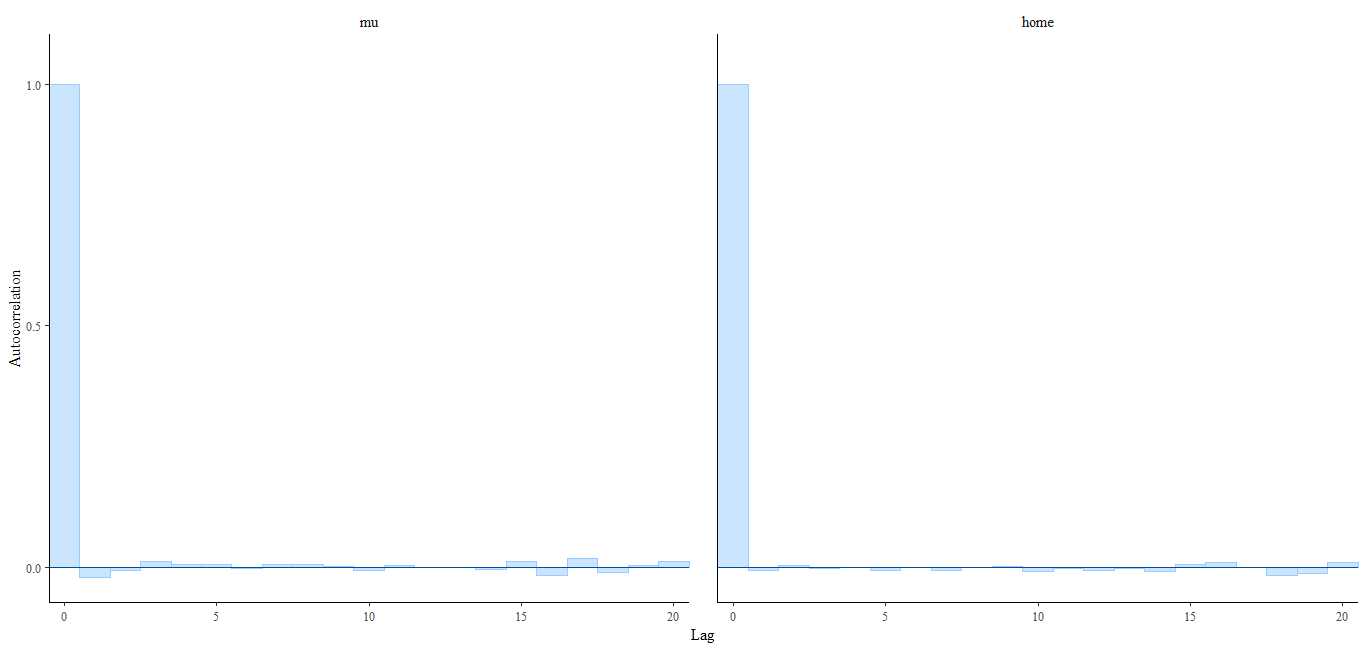}
		\caption{ACF plots for the constant parameter $\mu$ and the home effect of the ZDTS model.}
		\label{zdts_acf_mu_home_chains3}
		\end{center} 
	\end{figure} 
	
	\begin{figure}[h!]
		\includegraphics[width=1\textwidth,height=0.8\textheight,keepaspectratio]{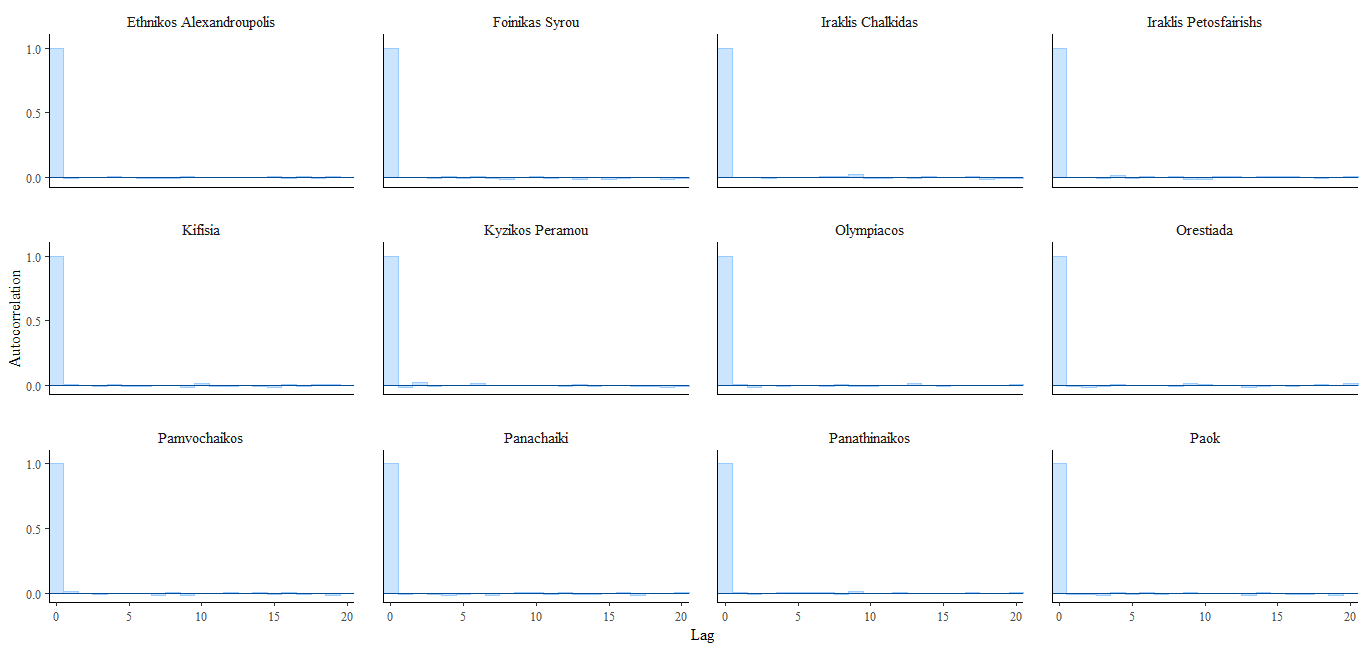}
		\caption{ACF plots for the attacking abilities parameters of the ZDTS model.}
		\label{zdts_acf_attack_chains3}
	\end{figure} 
	
	\begin{figure}[h!]
		\includegraphics[width=1\textwidth,height=0.8\textheight,keepaspectratio]{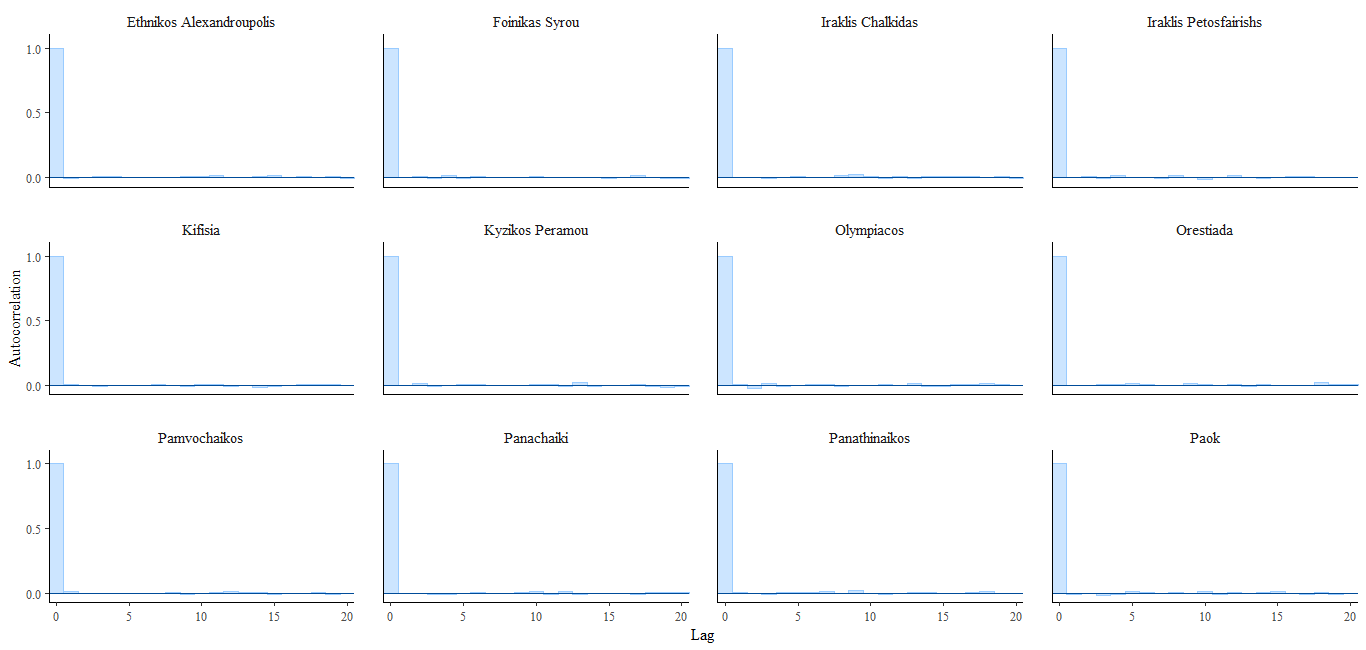}
		\caption{ACF plots for the defensive abilities parameters of the ZDTS model.}
		\label{zdts_acf_defense_chains3}
	\end{figure} 
	
	%---Ergodic
	
	\begin{figure}[h!]
		\includegraphics[width=1\textwidth,height=0.8\textheight,keepaspectratio]{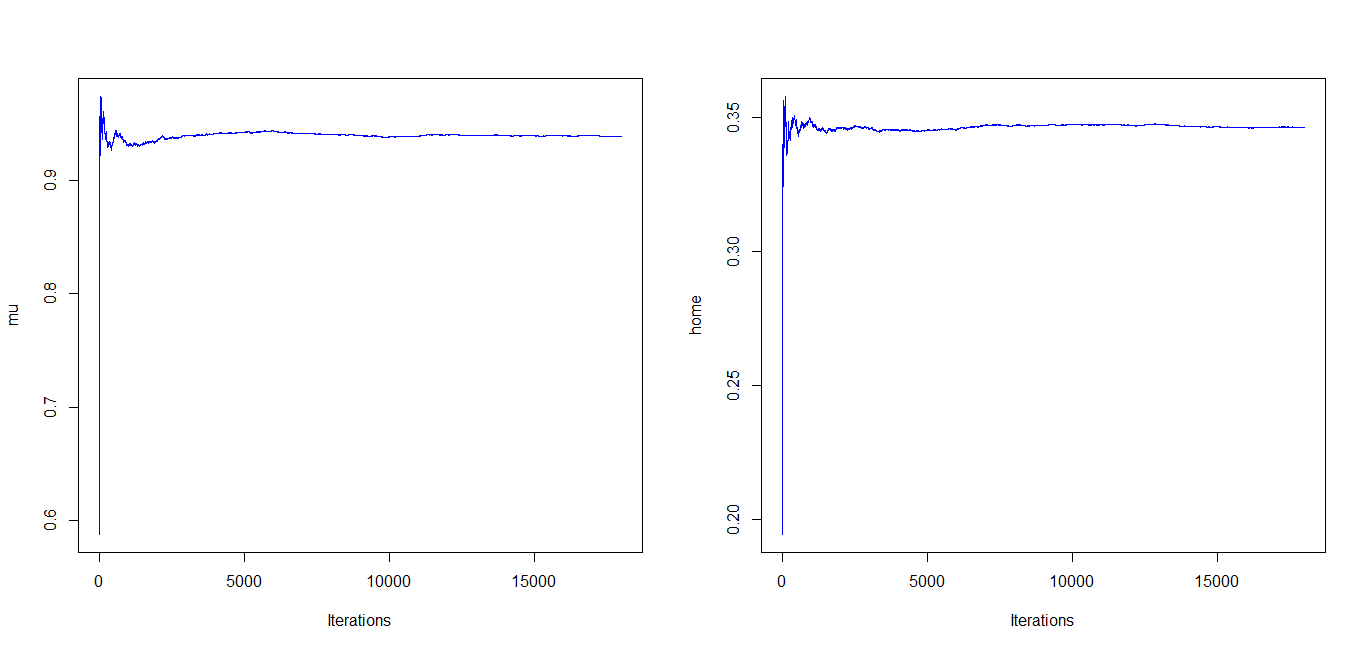}
		\caption{Ergodic plots for the constant parameter $\mu$ and the home effect of the ZDTS model.}
		\label{zdts_erg_mu_home_chains3}
	\end{figure} 
	
	\begin{figure}[h!]
		\includegraphics[width=1\textwidth,height=0.8\textheight,keepaspectratio]{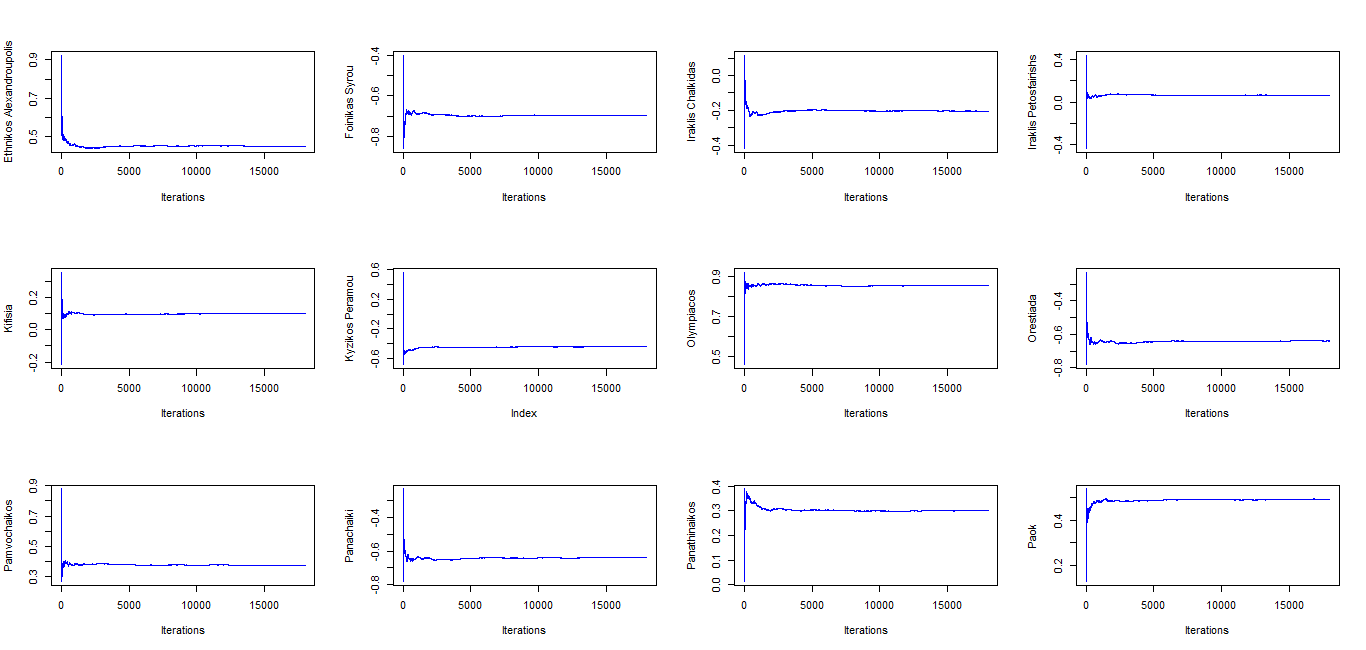}
		\caption{Ergodic plots for the attacking abilities parameters of the ZDTS model.}
		\label{zdts_erg_attack_chains3}
	\end{figure} 
	
	\begin{figure}[h!]
		\includegraphics[width=1\textwidth,height=0.8\textheight,keepaspectratio]{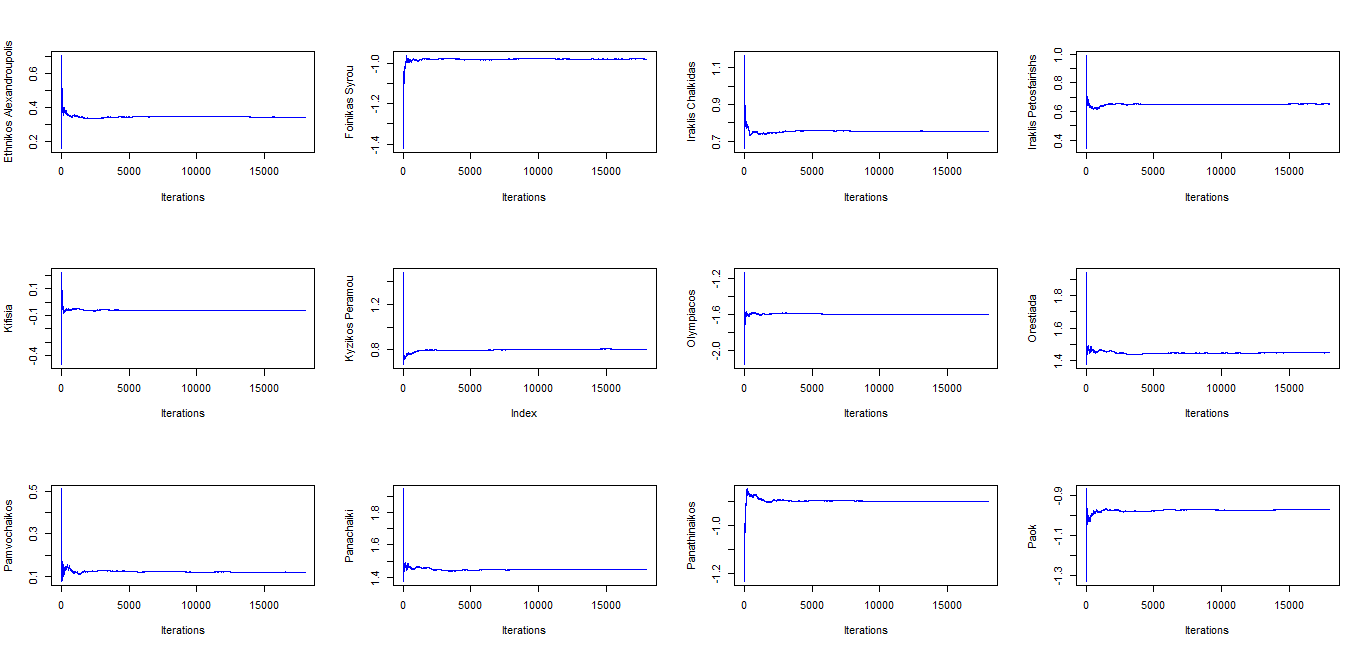}
		\caption{Ergodic plots for the defensive abilities parameters of the ZDTS model.}
		\label{zdts_erg_defense_chains3}
	\end{figure}

	%Trace

	\begin{figure}[h!]
		\includegraphics[width=1\textwidth,height=0.8\textheight,keepaspectratio]{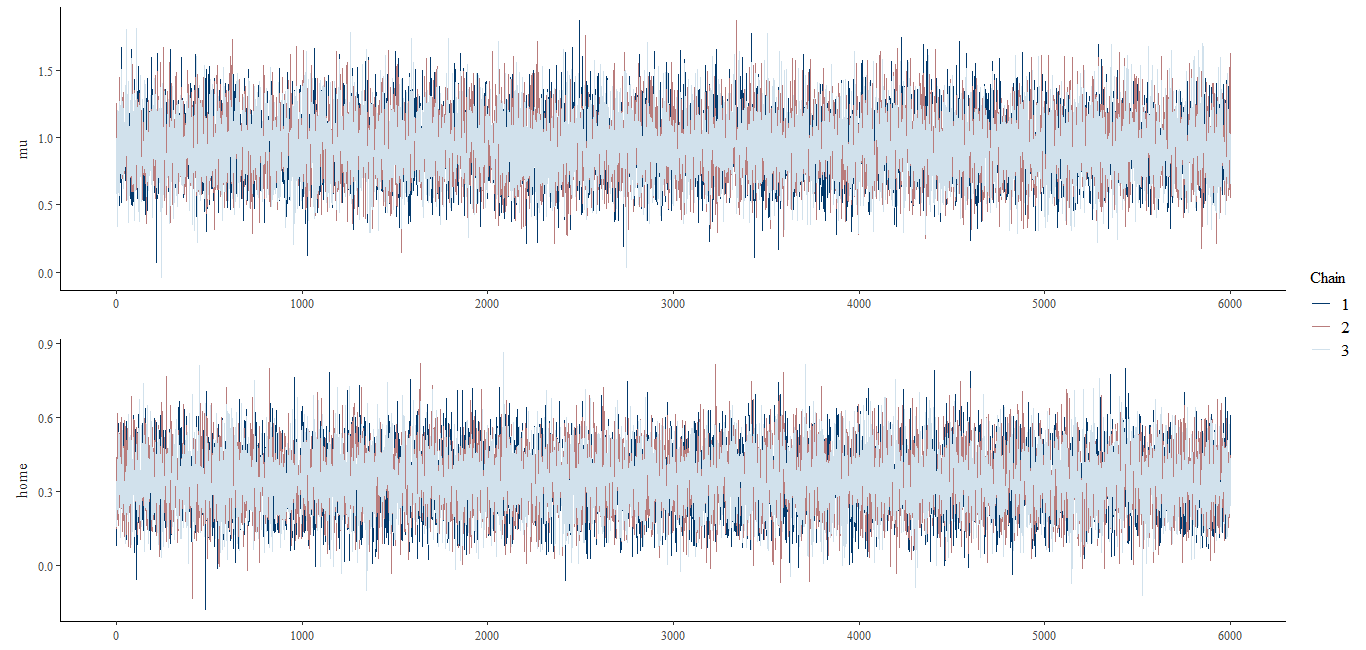}
		\caption{Trace plots for the constant parameter $\mu$ and the home effect of the ZDTS model.}
		\label{zdts_trace_mu_home_chains3}
	\end{figure}
	
	\begin{figure}[h!]
		\includegraphics[width=1\textwidth,height=0.8\textheight,keepaspectratio]{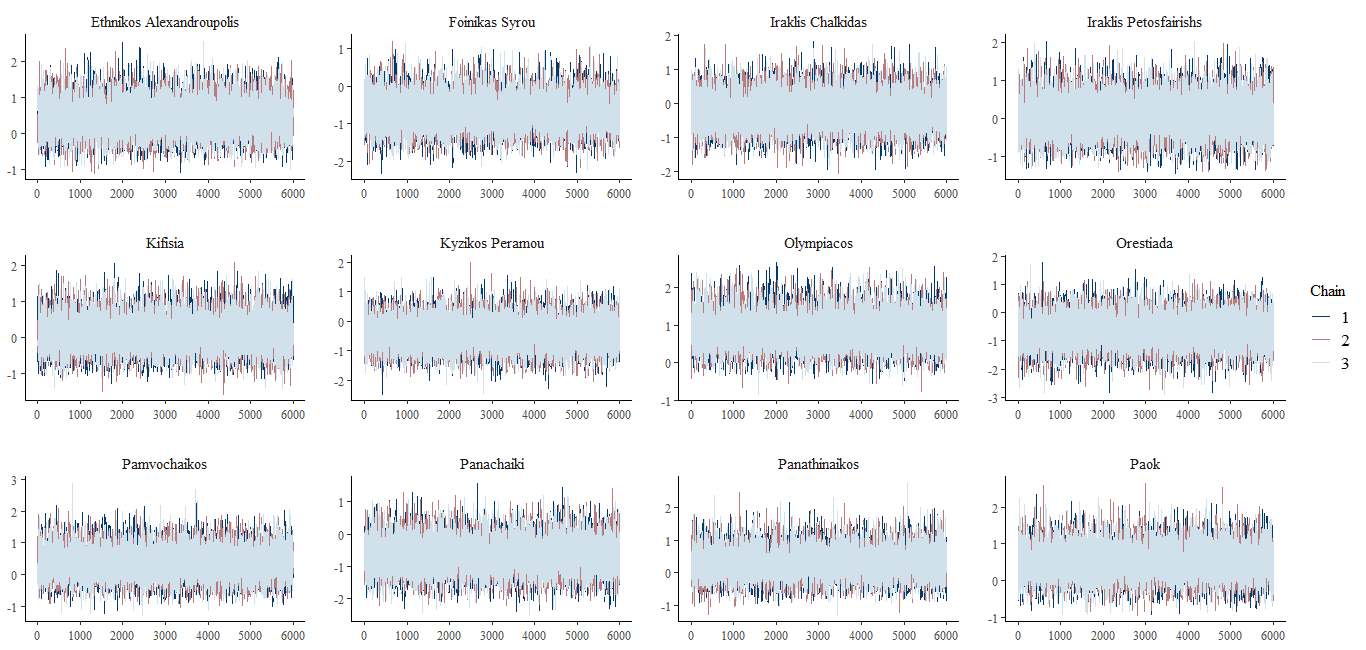}
		\caption{Trace plots for the attacking abilities parameters of the ZDTS model.}
		\label{zdts_trace_defense_chains3}
	\end{figure} 
	
	\begin{figure}[h!]
		\includegraphics[width=1\textwidth,height=0.8\textheight,keepaspectratio]{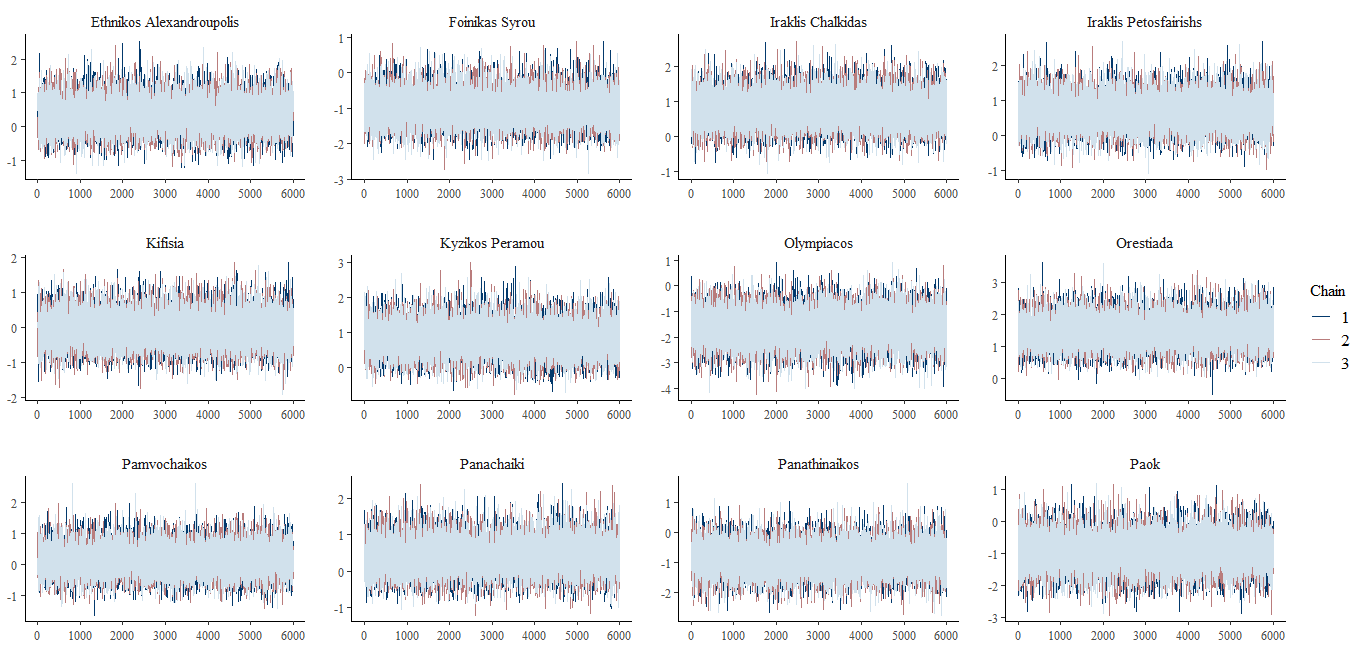}
		\caption{Trace plots for the defensive parameters of the ZDTS model.}
		\label{zdts_trace_attack_chains3}
	\end{figure}

	\clearpage
	
	\section{Additional posterior inference (Additional details for Section 4.1)}
	\label{appendix_C}
	
	Additional posterior results for Section 4.1 of the main article are provided in \ref{appendix_C} which is divided in three sections. 
	Appendix \ref{App_C1} presents summary tables of the posterior distribution of the parameters of the two   fitted models. 
	In Appendix \ref{App_C2}, you can find the density plots of the posterior distributions of the percentages for each possible value of the set-difference for a match between two teams of equal strength.  
	Appendix \ref{App_C3} presents the details of the approximate analysis which can be used to obtain an easier interpretation about the parameters of the ZDTS as a function of the expected  set-difference. 
	This analysis was added after a comment from a referee.

\subsection{Posterior summaries of model parameters} 
\label{App_C1}

\begin{table}[ht]
	\centering
	\begin{tabular}{lrr}
		\hline
		& \multicolumn{2}{c}{Posterior} \\ 
		Parameter & Mean & St.dev. \\ 
		\hline
		\hline 
		Constant Parameters & \\ 
		\hline 
		$c_1$ & -2.83 & 0.34 \\ 
		$c_2$ & -1.25 & 0.26 \\ 
		$c_3$ & -0.45 & 0.24 \\ 
		$c_4$ & 0.36 & 0.24 \\ 
		$c_5$ & 1.98 & 0.29 \\ 
		\hline 
		Ability Parameters & \\ 
		\hline 
		Ethnikos Alexandroupolis  & 0.16 & 0.39 \\ 
		Foinikas Syrou  & 0.67 & 0.37 \\ 
		Iraklis Chalkidas  & -1.09 & 0.39 \\ 
		Iraklis Petosfairishs  & -0.96 & 0.38 \\ 
		Kifisia  & 0.40 & 0.38 \\ 
		Kyzikos Peramou  & -1.60 & 0.43 \\ 
		Olympiacos  & 2.79 & 0.47 \\ 
		Orestiada  & -3.04 & 0.51 \\ 
		Pamvochaikos  & 0.42 & 0.39 \\ 
		Panachaiki  & -1.23 & 0.41 \\ 
		Panathinaikos  & 1.55 & 0.38 \\ 
		PAOK  & 1.94 & 0.40 \\ 
		\hline
		\multicolumn{3}{p{7cm}}{ \small \it \textbf{MCMC output details:} chains=3; 
			For each chain: total length=$16000$ iterations, Burnin/warmup iterations $B=2000$, thin=2, iterations kept $T=6000$; Total number of iterations finally used=$18000$ iterations ($3\times 6000$)}
	\end{tabular}
	\caption{Posterior summary statistics for the parameters of the ordered-multinomial model.} 
	\label{Ordered_Post.sum}
\end{table}

\begin{table}[ht]
	\centering
	\begin{tabular}{lrr}
		\hline
		 & \multicolumn{2}{c}{Posterior} \\ 
		Parameter & Mean & St.dev. \\ 
		\hline
		$\mu$ & 0.94 & 0.24 \\ 
		home & 0.35 & 0.12 \\ 
		\hline 
		Attacking Effects & \\ 
		\hline 
		Ethnikos Alexandroupolis  & 0.45 & 0.47 \\ 
		Foinikas Syrou  & -0.70 & 0.43 \\ 
		Iraklis Chalkidas  & -0.20 & 0.50 \\ 
		Iraklis Petosfairishs  & 0.07 & 0.48 \\ 
		Kifisia  & 0.10 & 0.46 \\ 
		Kyzikos Peramou  & -0.44 & 0.51 \\ 
		Olympiacos  & 0.85 & 0.44 \\ 
		Orestiada  & -0.64 & 0.56 \\ 
		Pamvochaikos  & 0.37 & 0.46 \\ 
		Panachaiki  & -0.66 & 0.51 \\ 
		Panathinaikos  & 0.30 & 0.44 \\ 
		PAOK  & 0.49 & 0.44 \\ 
		\hline 
		Defensive Effects & \\ 
		\hline 
		Ethnikos Alexandroupolis  & 0.34 & 0.48 \\ 
		Foinikas Syrou  & -0.98 & 0.45 \\ 
		Iraklis Chalkidas  & 0.75 & 0.47 \\ 
		Iraklis Petosfairishs  & 0.65 & 0.46 \\ 
		Kifisia  & -0.06 & 0.47 \\ 
		Kyzikos Peramou  & 0.81 & 0.45 \\ 
		Olympiacos  & -1.61 & 0.65 \\ 
		Orestiada  & 1.45 & 0.47 \\ 
		Pamvochaikos  & 0.12 & 0.47 \\ 
		Panachaiki  & 0.39 & 0.45 \\ 
		Panathinaikos  & -0.90 & 0.50 \\ 
		PAOK & -0.97 & 0.53 \\ 
		\hline
		\multicolumn{3}{p{7cm}}{ \small \it \textbf{MCMC output details:} chains=3; 
			For each chain: total length=$16000$ iterations, Burnin/warmup iterations $B=2000$, thin=2, iterations kept $T=6000$; Total number of iterations finally used=$18000$ iterations ($3\times 6000$)}
	\end{tabular}
	\caption{Posterior summary statistics for the parameters of the ZDTS model.} 
\label{ZDTS_Post.sum}
\end{table}

\clearpage
\newpage 

	\subsection{Posterior distributions of set-differences (Additional to Table 3)} 
	\label{App_C2}
	\begin{figure}[h!]
		\includegraphics[width=1\textwidth,height=0.8\textheight,keepaspectratio]{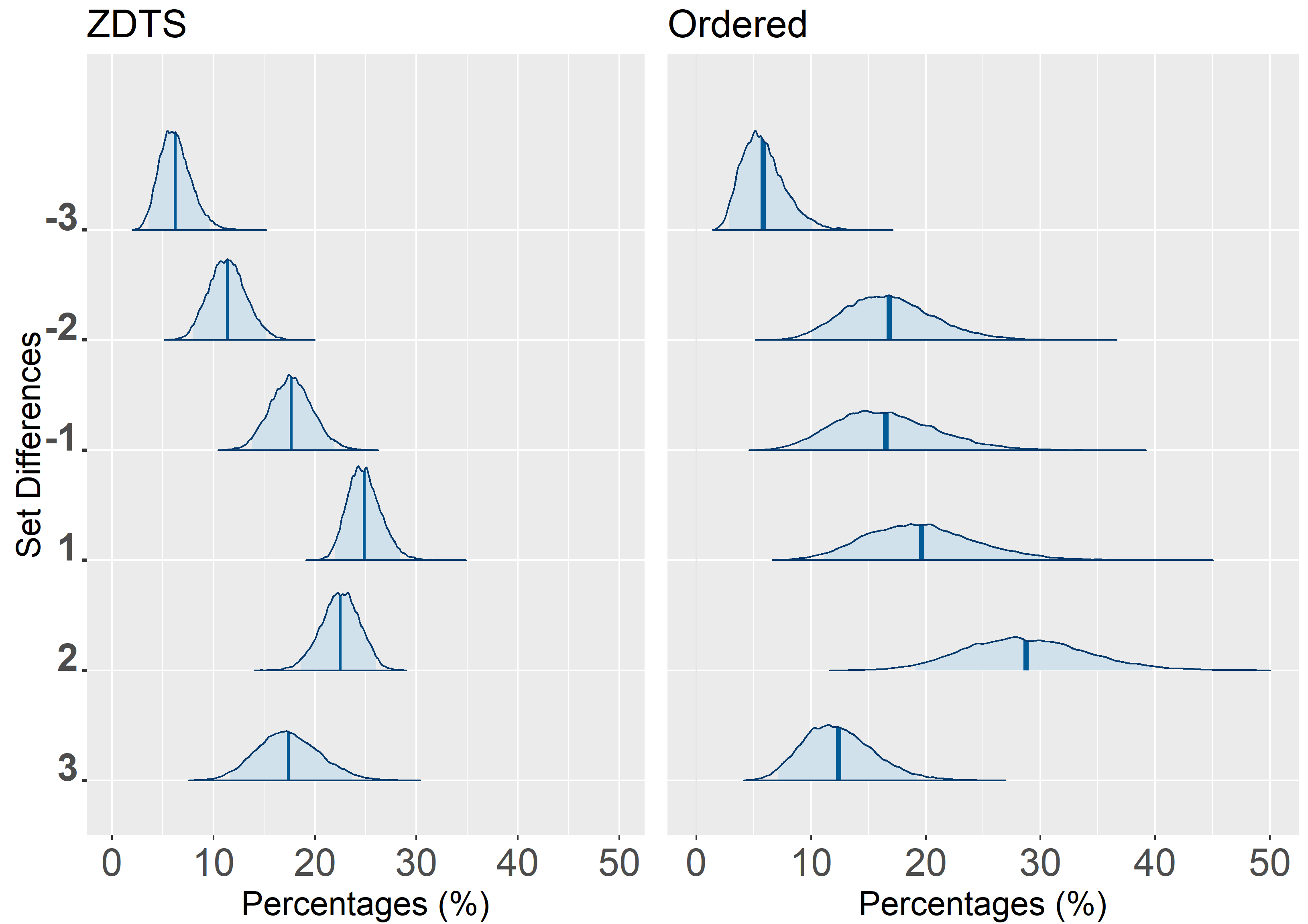}
		\caption{95\% Posterior densities of predicted percentages (\%) of set-differences between two equal strength teams based on the ZDTS and ordered-multinomial model, respectively.}
		\label{ord_zdts_probabilities_outcomes}
	\end{figure}

	\clearpage

	\subsection{The covariate effect as approximate proportional change of the shifted expected set-difference}
	\label{App_C3}

	Following a reviewer's comment about the effect of each covariate on the expected value of the ZDTS distribution, 
	we have fitted a series of models using the posterior values of $\lambda_1$ and $\lambda_2$ and the expected set-difference $E(Z_{ZDTS})$. 
	Here we present  results from the simple linear regression model with response the quantity $\log \big[ 3+E(Z_{ZDTS}) \big]$ and the difference between $\log \lambda_1$ and $\log \lambda_2$ as an explanatory variable. 
	Both these quantities  have been calculated in every MCMC iteration and for each observation. 
	This model was selected among other competitors we have fitted because it combines a satisfactory fit and a straightforward interpretation. 
	The approximating equation that this model induces is 
	\begin{equation}\label{zdts_sk_model}
	\log \big[ 3+E(Z_{ZDTS}) \big]=\alpha + \beta * \log\frac{\lambda_{1}}{\lambda_{2}} 
	\end{equation}
	where $\log(\lambda_{1})=b_{0}+b_{1}x$ and $\log(\lambda_{2})=b_{0}'+b_{2}x$ and $\lambda_{1}$, $\lambda_{2}$ are the parameters of the ZDTS model. 
	The quantity $ 3+E(Z_{ZDTS}) $  will be referred as the shifted expected set-difference (SESD). 
	
	If this model  holds, then the proportional change of the shifted expected set-difference, 
	$SESD(x)=3+E(Z_{ZDTS}|X=x)$, for a change of one unit in a covariate $X$ is given by 
	\begin{eqnarray} 
	\%\Delta \big[ E(Z_{ZDTS}) \big] 
	&=& \frac{SESD(x+1) - SESD(x)}{SESD(x)} 
	= \frac{E(Z_{ZDTS} | X=x+1) - E(Z_{ZDTS} | X=x)}{3+E(Z_{ZDTS}|X=x)} \nonumber \\ 
	&=&\big(e^{\beta(b_1-b_2)}-1\big) 
	\label{parameter_interpretation}
	\end{eqnarray} 
	
	This is simply obtained by the following steps/calculations: 
	\begin{enumerate}
		\item For the $\lambda_1$ when $X=x+1$ we have that $\log\big( \lambda_1(x+1) \big)=b_0+b_1(x+1) = b_0+b_1x+b_1= \log\big( \lambda_1(x) \big)+b_1$. 
		\item Similarly, for $\lambda_2$ we have that $\log\big( \lambda_2(x+1) \big)= \log\big( \lambda_2(x) \big)+b_2$. 
		\item The shifted expected set-difference when $X=x$ is given by $SESD(x)=3+E(Z_{ZDTS}|X=x) = e^{ \alpha + \beta \log\big( \lambda_1(x) \big) - \beta \log\big( \lambda_2(x) \big) }$. 
		\item The shifted expected set-difference when $X=x+1$ is given by 
		$$
		SESD(x+1)=3+E(Z_{ZDTS}|X=x+1) = e^{ \alpha + \beta \log\big( \lambda_1(x+1) \big) - \beta \log\big( \lambda_2(x+1) \big) }. 
		$$
		From Steps 1 and 2, we have that 
		by 
		\begin{eqnarray*} 
		SESD(x+1) &=&  e^{ \alpha + \beta \log\big( \lambda_1(x) \big)+\beta b_1 - \beta \log\big( \lambda_2(x) \big) - \beta b_2} \\
		&=&SESD(x)e^{ \beta (b_1-b_2) } \Leftrightarrow \\  
		\frac{SESD(x+1)-SESD(x) }{SESD(x)} &=& e^{ \beta (b_1-b_2)}-1. 
		\end{eqnarray*}

	\end{enumerate}		
	 
	The parameters of the ZDTS model are expressing proportional differences on $\lambda_1$ and $\lambda_2$. 
	If expression \eqref{zdts_sk_model} explains satisfactorily the  relationship between $E(Z_{ZDTS})$ and $\lambda_1, ~\lambda_2$, 
	then, via  expression \eqref{parameter_interpretation},  we can directly transform the parameters of the ZTDS model to a proportional change of the shifted expected set-difference. 
	To be more specific, the quantity $100 \times [ e^{ \beta (b_1-b_2)}-1]$ provides the percentage (proportional)   change  in the shifted expected set-difference (SESD) when $X$ increases by one unit. 
	Hence, the quantity $\beta$ estimated via \eqref{zdts_sk_model} plays the role of a correction factor providing the appropriate down-weight of the ZDTS parameters in order to be transformed in a proportional   change in terms of shifted expected set-difference.

\begin{figure}[h]
		\centering
		\includegraphics[width=0.8\textwidth,keepaspectratio]{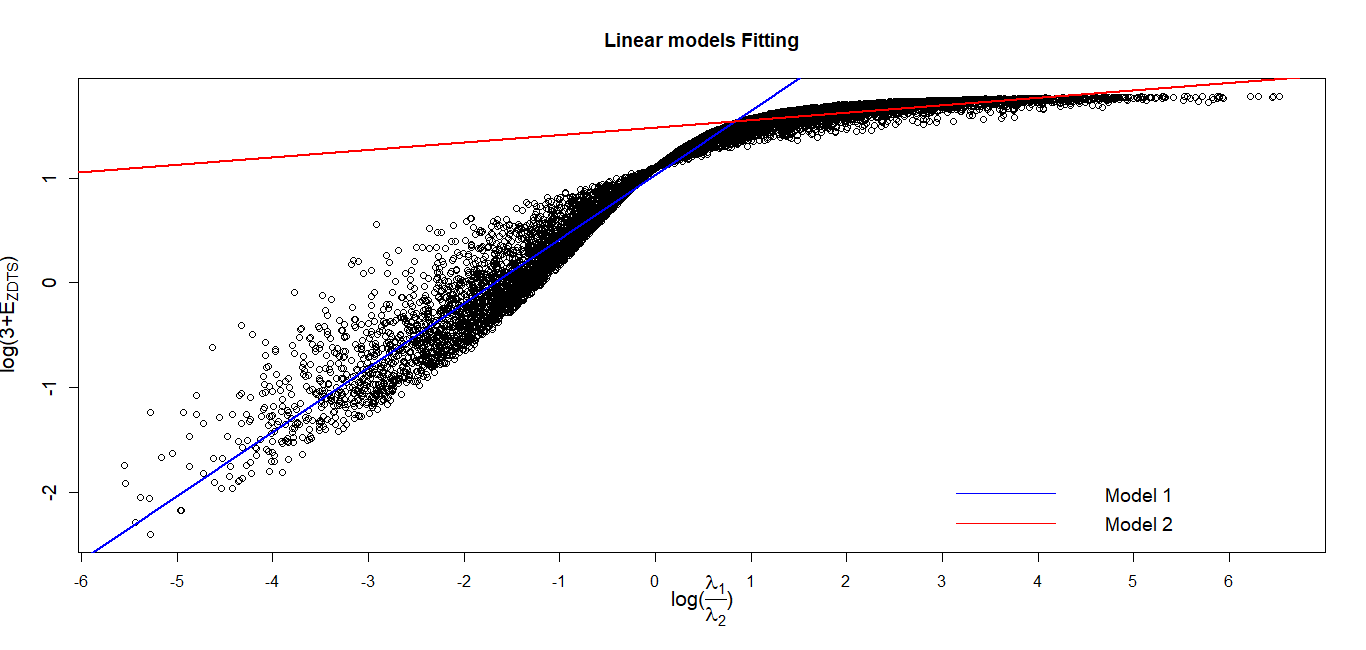}
		\caption{Fitting of the model \eqref{zdts_sk_model} for both cases of a) $\log\frac{\lambda_{1}}{\lambda_{2}} \leq 0.8$ (Model 1) and b) $\log\frac{\lambda_{1}}{\lambda_{2}} > 0.8$ (Model 2).}
		\label{zdts_sk_models_plot}
	\end{figure}

	Figure \ref{zdts_sk_models_plot} presents the connection between $\log SESD $ and $\log \frac{\lambda_{1}}{\lambda_{2}}$. 
	From this Figure, we fitted the simple model \eqref{zdts_sk_model} for two different cases: 
	$$
	\mbox{Case A: } \log\Big(\frac{\lambda_{1}}{\lambda_{2}}\Big) \leq C \mbox{~~~~and~~~~}  
	\mbox{Case B: } \log\Big(\frac{\lambda_{1}}{\lambda_{2}}\Big) > C; 
	$$
	where $C$ is an optimal cut-off point in terms of a model fit measure.

\begin{figure}[b!]
	\centering
	\includegraphics[width=0.8\textwidth, keepaspectratio]{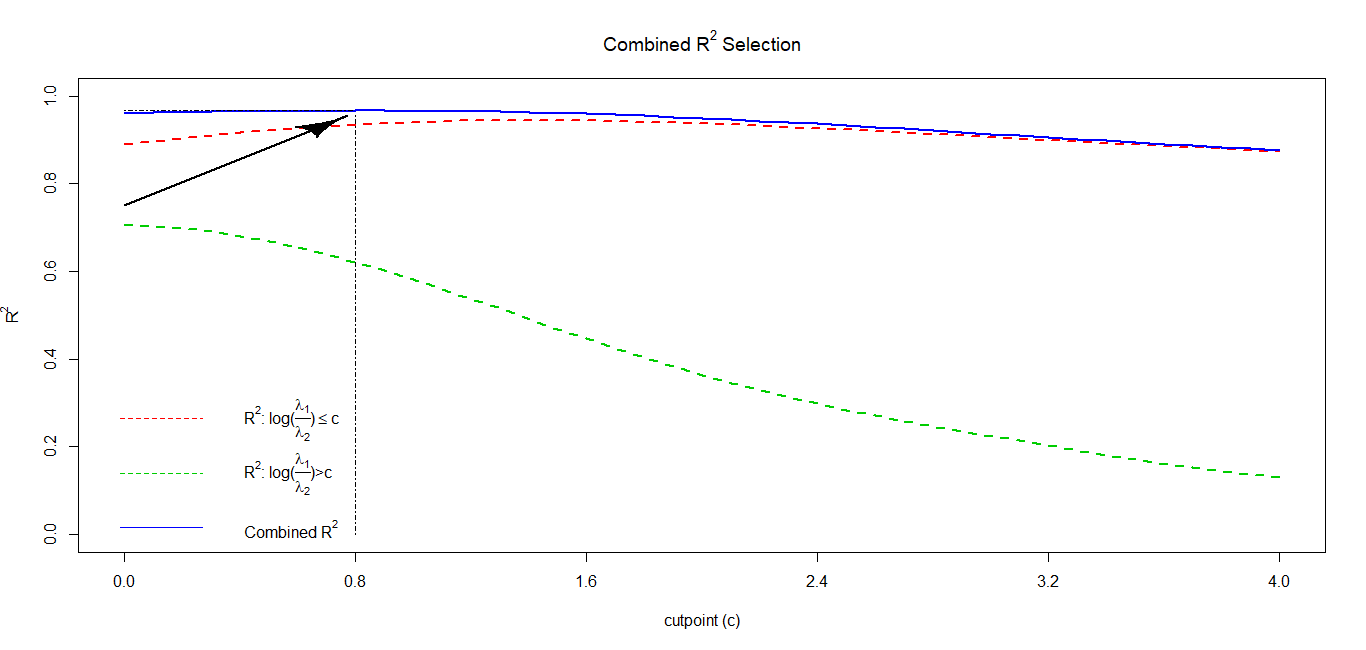}
	\caption{$R^{2}$ of Models 1, 2, along with the combined $R^{2}$.}
	\label{zdts_sk_combined_r2}
\end{figure}

	The cut-off point $C$ was selected  in such way that it maximizes an overall (combined) $R^2$ measure achieved by the combination of these two models. Figure \ref{zdts_sk_combined_r2} presents $R^{2}$ values of fitted models for 
$\log(\lambda_1/\lambda_2) \le C$ and $\log(\lambda_1/\lambda_2) > C$ as well as the combined $R^2$ for several choices of $C$.  We observe that the optimal cut-off point in terms of the overall (combined) $R^{2}$ is achieved for $C=0.8$.

	In each of the two fitted regression models, the interpretation is different related to the proportional change of $SESD$ when we increase a covariate by one unit in the  ZDTS model structure. 
	After fitting model \eqref{zdts_sk_model} for the posterior values of $\log(\lambda_1/\lambda_2) \le 0.8$ 
	and $\log(\lambda_1/\lambda_2) > 0.8$  we found $\widehat{\beta}=0.61$ and $\widehat{\beta}=0.07$, respectively. 
	For each case/model, the interpretation is given by: 
	\begin{enumerate}[label=(\alph*)]
		\item For $\log \frac{\lambda_{1}}{\lambda_{2}}  \leq 0.8$: the proportional change of $SESD$, when a covariate $X$ is increased by one unit, is equal to $e^{0.61 b_{1}}$ when only $\lambda_{1}$ is influenced by this covariate (i.e. $b_1\neq 0$ and $b_2=0$) while this change is equal to $e^{0.61 (b_1-b_2)}$ when $X$ is a covariate for both $\lambda_{1}$ and $\lambda_{2}$ (i.e. $b_1\neq 0$ and $b_2\neq 0$).
		\item For $\log\frac{\lambda_{1}}{\lambda_{2}} > 0.8$: 
		the proportional change of $SESD$, when a covariate $X$ is increased by one unit, is equal to $e^{0.07 b_{1}}$ when only $\lambda_{1}$ is influenced by this covariate (i.e. $b_1\neq 0$ and $b_2=0$) while this change is equal to $e^{0.07 (b_1-b_2)}$ when $X$ is a covariate for both $\lambda_{1}$ and $\lambda_{2}$ (i.e. $b_1\neq 0$ and $b_2\neq 0$).
	\end{enumerate}

	\section{Model Diagnostics} 
	Here we present additional density plots of the predictive measures presented in Sections 4.3 and 4.4 of the main  body  of the paper. 
	
	\subsection{In-Sample Diagnostics (Additional plots of Section 4.3, Table 6)}
	\begin{figure}[ht!]
		\centering
		\hspace{-1cm}
		\includegraphics[width=0.9\textwidth,keepaspectratio]{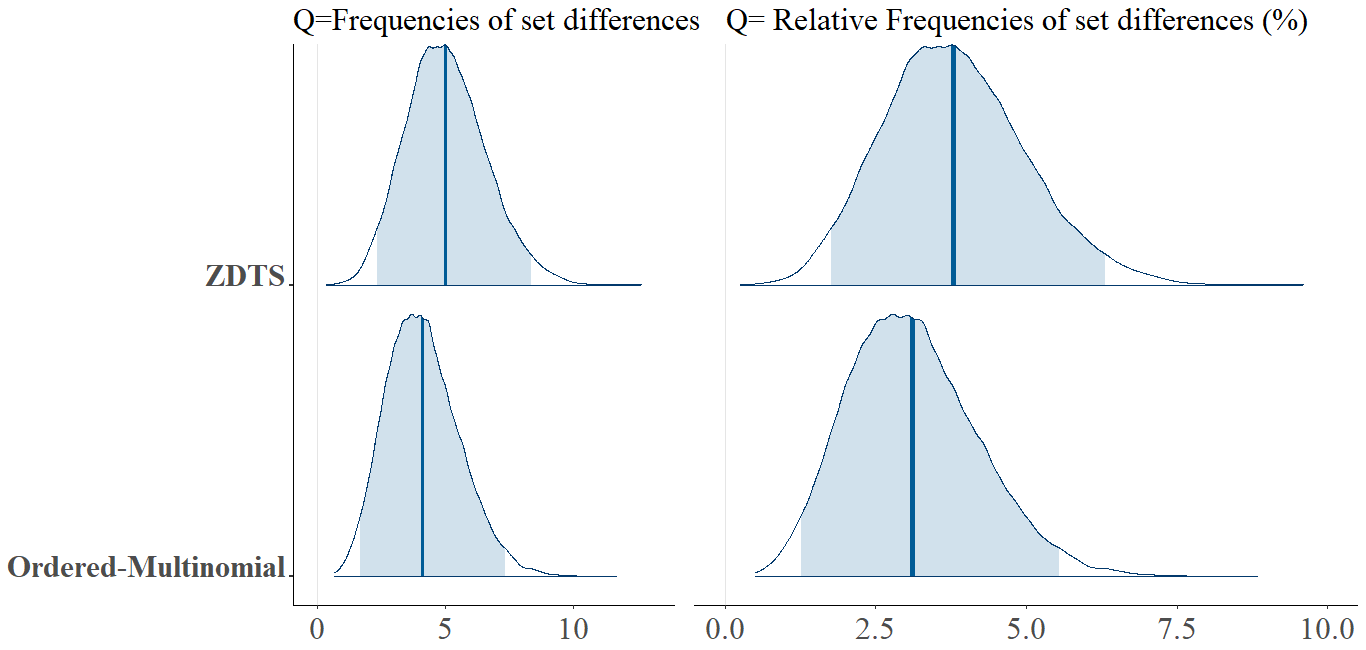}
		\centering
		\caption{Posterior distributions of MAD for frequencies and relative frequencies of set-differences,  $Q\in\{1,2\}$, for both fitted models (in-sample scenario based on the reproduced full league tables -- see Table 5 for summaries of the regenerated league tables and Table 6 for the summary statistics of MAD measures).}
	\end{figure}

\vspace{2cm} 
	
	\begin{figure}[ht!]
		\centering
		\hspace{-1cm}
		\includegraphics[width=0.9\textwidth,keepaspectratio]{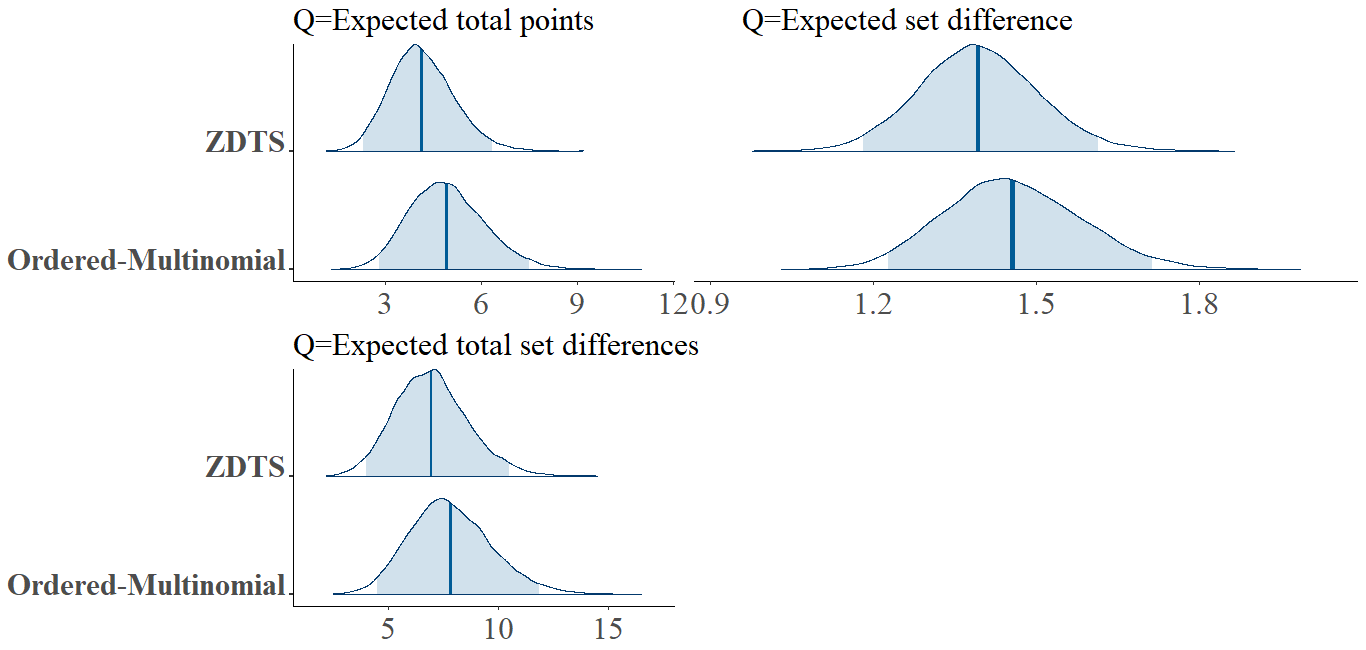}
		\centering
		\caption{Posterior distributions of MAD  for expected total points, expected differences and expected total set-difference, $Q\in\{3,4,5\}$,  for both fitted models (in-sample scenario based on the reproduced full league tables -- see Table 5 for summaries of the regenerated league tables and Table 6 for the summary statistics of MAD measures).}
	\end{figure}
	
	%\clearpage
	
	\subsection{Out-of-Sample Diagnostics (Additional plots of Section 4.4, Table 7)}
	\begin{figure}[ht!]
		\centering
		\hspace{-1cm}
		\includegraphics[width=0.9\textwidth,keepaspectratio]{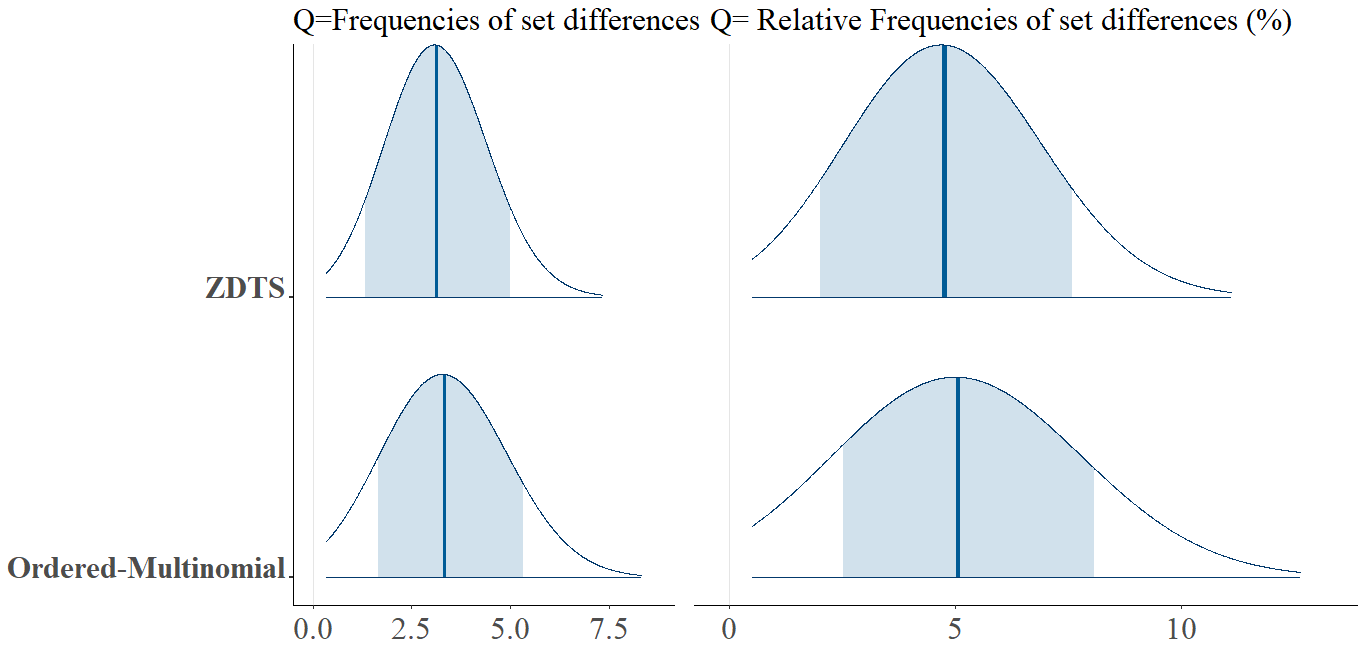}
		\centering
		\caption{Posterior distributions of MAD measures for frequencies and relative frequencies of set-differences,  $Q\in\{1,2\}$, for both fitted models (mid-season out-of-sample scenario based on the reproduced league tables -- see Table 6 for the summary statistics of MAD measures).}
	\end{figure}

\vspace{2cm}
	
	\begin{figure}[ht!]
		\centering
		\hspace{-1cm}
		\includegraphics[width=0.9\textwidth,keepaspectratio]{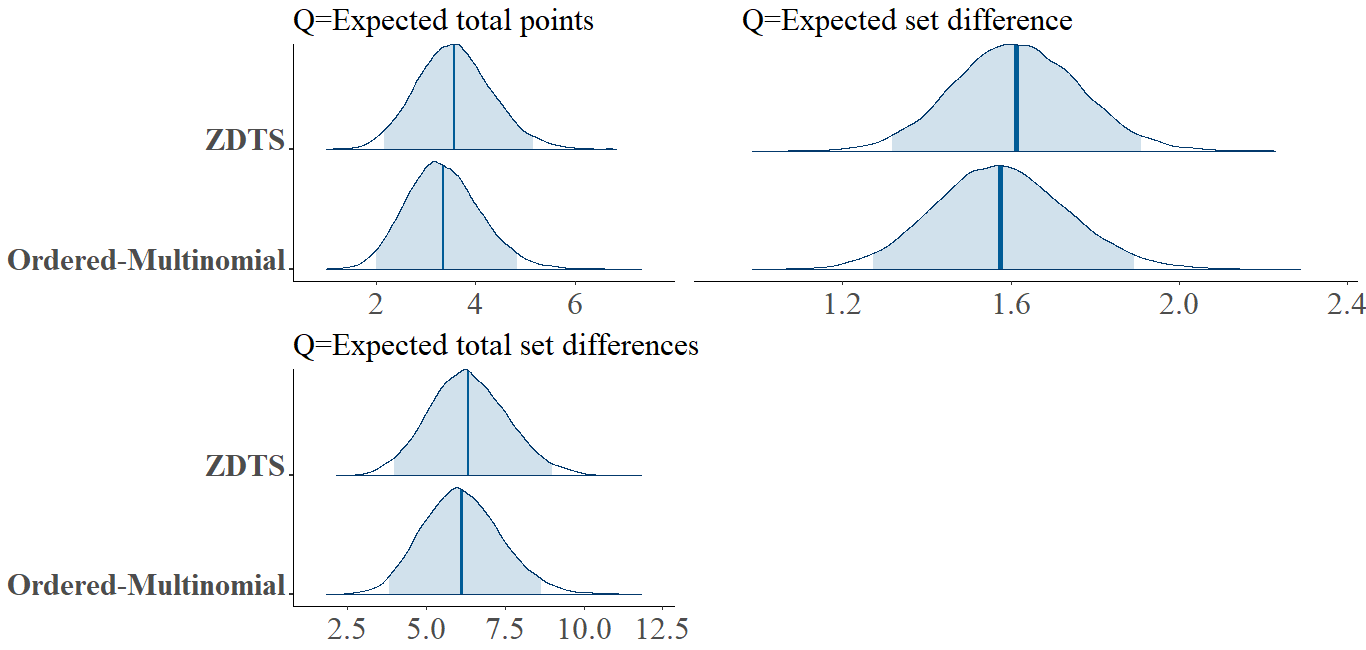}
		\centering
		\caption{Posterior distributions of MAD measures for expected total points, expected differences and expected total set-differences, $Q\in\{3,4,5\}$,  for both fitted models (mid-season out-of-sample scenario based on the reproduced league tables -- see Table 6 for the summary statistics of MAD measures).}
	\end{figure}